\documentclass{desyproc}

\begin{document}
%------------------------------------
\title{Charmonium %and heavy mesons 
	in a hot, dense medium}

%for single authors the superscripts are optional
\author{{\slshape David Blaschke$^{1,2}$\\[1ex]
$^1$Institute for Theoretical Physics, University of Wroc{\l}aw, 
%Max-Born pl. 9, 
50-204 Wroc{\l}aw, Poland\\
$^2$Bogoliubov Laboratory for Theoretical Physics, JINR, 141980 Dubna, Russia}
}
% if the proceedings are available online (e.g. at Indico)
% please enter the contribution ID or file_name below for the DOI
%\contribID{32}
\contribID{blaschke\_david}

% TO THE CONFERENCE EDITORS: 
% please update the following information      
% before sending the template to the authors
% \confID{800}  % if the conference is on Indico uncomment this line
\desyproc{DESY-PROC-2009-07}
\acronym{HQP08} % if you want the Acronym in the page footer uncomment this line
\doi  % if there is an online version we will register DOIs

\maketitle

\begin{abstract}
In this lecture we apply a thermodynamic Green function 
formalism developed in the context of nonrelativistic plasma physics for the
case of heavy quarkonia states in strongly correlated quark matter.
Besides the traditional explanation of charmonium suppresion by 
Debye screening of the strong interaction, we discuss further effects of
relevance when heavy quarkonia states propagate in a medium where strong 
correlations persist in the form of hadronic resonances.
These effects may be absorbed in the definition of a plasma Hamiltonian, which
was the main result of this work. This plasma Hamiltonian governs the 
in-medium modification of the bound state energy levels as well as the lowering
of the continuum edge which leads not only to the traditional Mott effect for
the dissociation of bound states in a plasma, but can also be applied for a 
consistent calculation of the in-medium modification of quarkonium 
dissociation rates. 
\end{abstract}

\section{Introduction}

In developing a theoretical approach to heavy quarkonia as messengers of the
deconfinement/ hadronization transition of a quark-gluon plasma formed
in a heavy-ion collision, we should aim at a unifying description
where hadrons appear as bound states (clusters) of quarks and gluons.
The situation is analoguous to the problem of two-particle states in QED
plasmas where a well-developed theory in the framework of the Green function
technique exists.
These methods have been widely developed for the case of the hydrogen plasma,
where the electrons and protons as the elementary constituents can form
hydrogen atoms as bound states of the attractive Coulomb interaction.
The problem is tractable analytically for the isolated two-particle system,
with a discrete energy spectrum of bound states and a continuous spectrum of
scattering states. Higher complexes, such as molecular hydrogen can also be
formed.
%%%%%%%%%%

In a many-particle system, the problem of bound state formation needs to
account for medium effects. They give contributions to a plasma Hamiltonian
%beyond the isolated two-body problem
\begin{equation}
\label{plasmaH}
H^{\rm pl} = H^{\rm Hartree}+ H^{\rm Fock}+ H^{\rm Pauli}+ H^{\rm MW}
+ H^{\rm Debye}+ H^{\rm pp}+ H^{\rm vdW}+ \dots ,
\end{equation}
where the first three effects, the Hartree- and Fock energies of one-particle
states and the Pauli blocking for the two-particle states,  are of
first order with respect to the interaction and determine the mean-field
approximation.
The following two terms of the plasma Hamiltonian are the Montroll-Ward term
giving the dynamical screening of the interaction in the self-energy, and the
dynamical screening (Debye) of the interaction between the bound particles.
These contributions are related to the polarization function and are of
particular interest for plasmas due to the long-range character of the Coulomb
interaction.
In a consistent description, both terms should be treated simultaneously.
The last two contributions to the plasma Hamiltonian are of second order with
respect to the fugacity: the polarization potential, describing the
interaction of a bound state with free charge carriers, and the van der
Waals interaction, accounting for the influence of correlations (including 
bound states) in the medium on the two-particle system under consideration, 
see \cite{Redmer:1997,Ebeling:1986}.

Approximations to medium effects in the self-energy and the effective
interaction kernel have to be made in a consistent way, resulting in
predictions for the modification of one-particle and two-particle states.
On this basis, the kinetics of bound state formation and breakup processes
can be described, establishing the ionization equilibrium under given
thermodynamical conditions \cite{Schlanges:1988}.
Coulomb systems similar to the hydrogen plasma are electron-hole plasmas in
semiconductors \cite{Zimmermann:1978}, where excitons and biexitons play the
role of the atoms and molecules.
Other systems which have been widely studied are expanded fluids
like alkali plasmas or noble gas plasmas, see \cite{Redmer:1997} and
references therein. Applications of the plasma physics concepts for cluster
formation and Mott effect to the rather short-ranged strong interactions
have been given, e.g., in \cite{Ropke:1982,Ropke:1983} for nuclear matter and
in \cite{Horowitz:1985tx,Ropke:1986qs} for quark matter.

In this subsection, we want to discuss basic insights from these
investigations of bound state formation in plasmas, as far as they can
concern our discussion of heavy quarkonia formation in hot and dense matter.
Before going more into the details, let us mention them.
Bound state properties remain rather inert to changes of the medium since the
self-energy and interaction effects partially compensate each other in lowest 
order of density. 
Also, the smaller size of the bound states matters in this respect.
The compensation does not hold for continuum states,
being influenced by self-energy effects only, so that a lowering of the
in-medium ionization threshold must occur which leads to a strong enhancement
of the
rate coefficients for  bound-free transitions and to a sequential ``melting''
of different bound state excitation levels into the continuum of scattering
states at corresponding critical plasma parameters
(Mott effect \cite{Mott:1968}), until even the ground state becomes unbound.

The theory of strongly coupled plasmas has been developed also for strong
nonideality, where the formation of clusters in the medium need to be taken
into account. This situation is similar to that of a hadronizing quark-gluon
plasma and we will therefore refer to cluster expansion techniques as the
theoretical basis.

\section{Bethe-Salpeter equation and plasma Hamiltonian}

\begin{figure}[!ht]
\includegraphics[width=0.45\textwidth,angle=0]{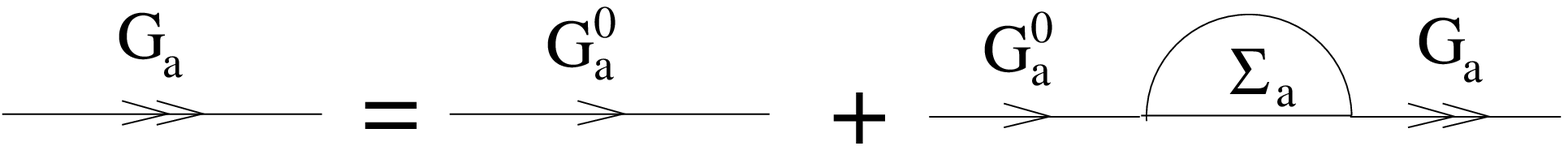}
~~;~~
%\hspace{5mm}
\includegraphics[width=0.45\textwidth,angle=0]{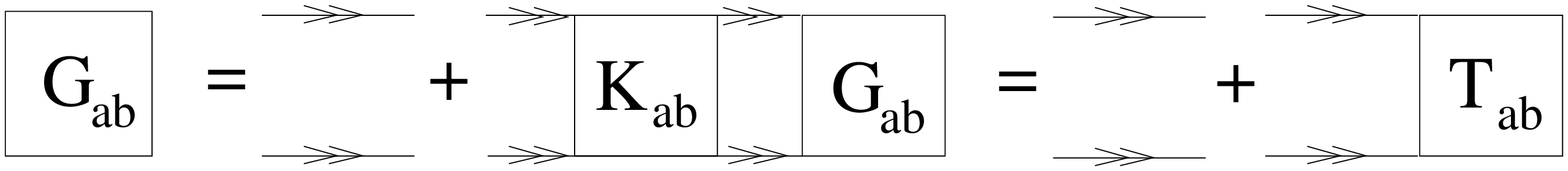}
%\end{tabular}
\caption{The two-particle problem in the medium. Dyson equation
(left) and Bethe-Salpeter equation (right) need to be solved in consistent
(conserving) approximations for self-energy ($\Sigma$) and interaction kernel
($K$).
\label{fig:BSE-Dyson} }
\end{figure}

The most systematic approach to the description of bound states in plasmas
uses the Bethe-Salpeter equation (BSE) for the thermodynamic (Matsubara-)
two-particle Green function 
\begin{equation}
\label{BSE}
G_{ab} = G_{ab}^0 + G_{ab}^0~K_{ab}~G_{ab} 
= G_{ab}^0 + G_{ab}^0~T_{ab}~G_{ab}^0~,
\end{equation}
which is equivalent to the use of the two-particle $T$-matrix $T_{ab}$ and
has to be solved in conjunction with the Dyson equation for the full
one-particle Green function,
\begin{equation}
\label{Dyson}
G_a = G_a^{0} + G_a^{0} \Sigma_a G_a~,
\end{equation}
defined by the dynamical self-energy $\Sigma_a(p,\omega)$ and the
free one-particle Green function
$G_a^{0}(p,\omega)=[\omega - \varepsilon_a(p)]^{-1}$
for a particle of species $a$ with the dispersion relation 
$\varepsilon_a(p)=\sqrt{p^2+m^2_a}\approx m_a+p^2/(2m_a)$,
see Fig. \ref{fig:BSE-Dyson}.

The BSE contains all information about the spectrum of two-particle bound as
well as scattering states in the plasma.
A proper formulation of the plasma effects on the two-particle spectrum
is essential to understand why bound and scattering states are influenced in a
different way by the surrounding medium, leading to the Mott-effect for
bound states. 
We give here the essence of a detailed discussion to be found in
Ref. \cite{Ebeling:1986}.

The homogeneous BSE associated with (\ref{BSE}) can be given the form of an 
effective Schr\"odinger equation for the wave function $\psi_{ab}(p_1,p_2,z)$
of two-particle states in the medium
\cite{Zimmermann:1978}
\begin{eqnarray}
\label{wave-eq}
 &&\sum_q\left\{\left[\varepsilon_a(p_1)+ \varepsilon_b(p_2) - z \right] 
%\psi_{ab}(p_1p_2,z)
\delta_{q,0}
-V_{ab}(q)\right\} \psi_{ab}(p_1+q,p_2-q,z)=
\nonumber\\
&&\hspace{3cm} 
=\sum_q H_{ab}^{\rm pl}(p_1,p_2,q,z) \psi_{ab}(p_1+q,p_2-q,z),
\end{eqnarray}
where $a,b$ denote a pair of particles with 3-momenta $p_1$ and $p_2$ which
transfer a 3-momentum $2q$ in their free-space interaction $V_{ab}(q)$ and 
$z$ is a complex two-particle energy variable.  
The in-medium effects described by (\ref{BSE}) have been singled out in the 
definition of a plasma Hamiltonian, containing all modifications beyond the 
two-body problem in free space \cite{Ebeling:1986,Zimmermann:1978}
\begin{eqnarray}
\label{plasma-h}
&&H_{ab}^{\rm pl}(p_1,p_2,q,z)=
\underbrace{V_{ab}(q)\left[N_{ab}(p_1,p_2)-1\right]}_{\rm (i)~Pauli~blocking}~
-\underbrace{\sum_{q'} V_{ab}(q')\left[N_{ab}(p_1+q',p_2-q')-1\right]
}_{\rm (ii)~Exchange~self-energy}\delta_{q,0}
\nonumber\\
&&+\underbrace{\Delta V_{ab}(p_1,p_2,q,z)N_{ab}(p_1,p_2)}_{\rm (iii)~ 
Dynamically~screened~potential}
-\underbrace{\sum_{q'} 
\Delta V_{ab}(p_1,p_2,q',z)N_{ab}(p_1+q',p_2-q')
}_{\rm (iv)~Dynamical~self-energy}\delta_{q,0}~.
\end{eqnarray}
Here, $\Delta V_{ab}(p_1,p_2,q,z)=K_{ab}(p_1,p_2,q,z)-V_{ab}(q)$
stands for the in-medium modification of the bare interaction potential 
to a dynamically screened interaction kernel $K_{ab}(p_1,p_2,q,z)$.
The effects of phase space occupation
are encoded in the function $N_{ab}(p_1, p_2)$, which for the case of an 
uncorrelated fermionic medium takes the form of the Pauli blocking factor
$N_{ab}(p_1, p_2)=1-f_a(p_1)-f_b(p_2)$, where 
$f_a(p)=\{\exp[(\varepsilon_a(p)-\mu_a)/T]+1\}^{-1}$ is the Fermi distribution 
and $\mu_a$ the chemical potentential of the species $a$. 
Eq. (\ref{wave-eq}) is a generalization of the two-particle Schr\"odinger 
equation, where on the left-hand side the isolated two-particle problem is 
described while many-body effects due to the surrounding medium are given on 
the right-hand side.  
The in-medium effects named in the plasma Hamiltonian (\ref{plasmaH}) can be 
obtained from the one derived in the Bethe-Salpeter approach (\ref{plasma-h}) 
upon proper choice of the interaction kernel $K_{ab}$ so that 
Eq.~(\ref{plasmaH}) appears a sa special case of Eq.~(\ref{plasma-h}).

The influence of the plasma Hamiltonian on the spectrum of bound and scattering
states can be qualitatively discussed in perturbation theory.
Since bound states are localized in coordinate space, their momentum space wave
functions extend over a finite range $\Lambda$ and we may assume them to be 
$q$-independent: 
$\psi_{ab}(p_1+q,p_2-q,z=E_{nP})\approx \psi_{ab}(p_1,p_2,z=E_{nP})$ for small 
momentum transfer $q < \Lambda$ and to vanish otherwise. 
Assuming further a flat momentum dependence of the Pauli blocking factors  
$N_{ab}(p_1+q, p_2-q) \approx N_{ab}(p_1, p_2)$ for small $q$ where the 
interaction is strong, we obtain a cancellation of the Pauli blocking term (i) 
by the exchange self energy (ii) and of the dynamically screened potential 
(iii) by the dynamical self-energy (iv). 
Therefore, the bound states remain largely unmodified by medium effects. 
For scattering states which are extended in coordinate space and can be 
represented by a delta function in momentum space, the above cancellations do 
not apply and a shift of the two-particle continuum threshold results.
For this mechanism to work it is important that approximation schemes for the 
self-energy and the interaction kernel have to be consistent as, e.g., in the 
conserving scheme of $\Phi$-derivable theories \cite{Baym:1962sx}.

Summarizing the discussion of the plasma Hamiltonian: bound state energies 
remain unshifted to lowest order in the charge carrier density while the 
threshold for the continuum of scattering states is lowered. 
The intersection points of bound state energies and continuum threshold define 
the Mott densities (and temperatures) for bound state dissociation.

When applying this approach to heavy quarkonia in a medium where heavy quarks 
(either free or bound in heavy hadrons) are rare, then $N_{ab}=1$ so that both,
(i) and (ii) can be safely neglected.
The effects (iii) and (iv) stem from the dynamical coupling of the
%constituents of the 
two-particle state to collective excitations (plasmons) in the medium.
In the screened potential approximation, the interaction kernel
is represented by
$V^S_{ab}(p_1p_2,q,z)=
V^S_{ab}(q,z)\delta_{P,p_1+p_2}\delta_{2q,p_1-p_2}$ with
\begin{equation}
\label{V_S}
V^S_{ab}(q,z)=V_{ab}(q) + V_{ab}(q)\Pi_{ab}(q,z)V^S_{ab}(q,z)
=V_{ab}(q)[1-\Pi_{ab}(q,z)V_{ab}(q)]^{-1}~,
\end{equation}
with the total momentum $P$ and the momentum transfer $2q$ in the two-particle 
system. 
The most frequently used approximation for the here introduced polarization 
function $\Pi_{ab}(q,z)$, or for the equivalent dielectric function
$\varepsilon_{ab}(q,z)=1-\Pi_{ab}(q,z)V_{ab}(q)$, 
is the random phase approximation (RPA).
In the next two paragraphs we discuss the static, long wavelength limit of
the RPA and its generalization for a clustered medium.

\paragraph{Example 1: statically screened Coulomb potential.}

The systematic account of the modification of the interaction potential
between charged particles $a$ and $b$ by polarization of the medium is taken
into account in the dynamical polarization function $\Pi_{ab}(q,z)$, which in
RPA reads \cite{Ebeling:1986}
\begin{equation}
\Pi_{ab}^{\rm RPA}(q,z)=2 \delta_{ab} \int \frac{d^3p}{(2\pi)^3}
\frac{f_a(E_{p}^a)-f_a(E_{p-q}^a)}{E_{p}^a-E_{p-q}^a-z}~.
\end{equation}
For the Coulomb interaction, corresponding to the exchange of a massless
vector boson, the potential is obtained from the longitudinal propagator
in the Coulomb gauge is $V_{ab}(q)=e_a e_b/q^2$. 
For a recent discussion in the context of heavy quark correlators and 
potentials see, e.g., \cite{Brambilla:2008cx,Beraudo:2007ky}.
Due to the large masses of the constitutents in the heavy quarkonium case, one 
may use a Born-\-Oppenheimer expansion to replace the 
dynamically screened interaction by its static ($z=0$) and long-wavelength
($q\to 0$) limit.
For nondegenerate systems the distribution functions are Boltzmann 
distributions and their difference can be expanded as
\begin{equation}
f_a(E_{p}^a)-f_a(E_{p-q}^a)=
{\rm e}^{-E_p^a/T}\left(1- {\rm e}^{-(E_{p-q}^a-E_p^a)/T}\right)
\approx - f_a(E_{p}^a) (E_{p}^a-E_{p-q}^a-z)/T~,
\end{equation}
so that the energy denominator gets compensated and the polarization function
becomes 
\begin{equation}
\Pi_{ab}^{\rm RPA}(q,z)=-2 \frac{\delta_{ab}}{T} 
\int \frac{d^3p}{(2\pi)^3}f_a(E_{p}^a)=-\delta_{ab} \frac{n_a(T)}{T} ~.
\end{equation}
The corresponding dielectric function $\varepsilon^{\rm RPA}_{ab}(q,\omega)$
takes the form
\begin{equation}
\lim_{q\to 0}\varepsilon^{\rm RPA}(q,0)=1+\frac{\mu_D^2}{q^2}~,
~~~\mu_D^2 = \frac{1}{T}\sum_a e_a^2 n_a(T)~.
\end{equation}
The screened Coulomb potential in this approximation is therefore 
$V_{ab}^S(q)=V_{ab}(q)/\varepsilon^{\rm RPA}(q,0)=e_a e_b/(q^2+\mu_D^2)$.
%%%%%%%%%%%%
In this ``classical'' example  of the statically screened
Coulomb interaction, the contribution to the plasma Hamiltonian is real and
in coordinate representation it is given by
\begin{equation}
\label{V-eff}
\Delta V_{ab}(r) = -\frac{\alpha}{r}({\rm e}^{-\mu_D r}-1)
\approx \alpha \mu_D - \frac{\alpha}{2}\mu_D^2 r~,
\end{equation}
where $\alpha=e^2/(4\pi)$ is the fine structure constant.
For the change in the Hartree self-energy of one-particle states
due to Debye screening we can perform an estimate in momentum space
\begin{equation}
\label{Delta-SE}
\Sigma_{a} = \frac{4 \pi \alpha}{(2s_a+1)} \int\frac{d^3q}{(2\pi)^3}
\left[ \frac{1}{q^2+\mu_D^2} - \frac{1}{q^2}\right]f_a(E_q^a)
\approx -\frac{\alpha\mu_D^2}{\pi}\int_0^\infty \frac{dq}{q^2+\mu_D^2}
=-\frac{\alpha\mu_D}{2}~.
\end{equation}
This entails that to lowest order in the density the shift of the one-particle energies (continuum edge of unbound states) $\Sigma_a + \Sigma_b = - \alpha \mu_D$ compensates the contribution due to the screening of the interaction (\ref{V-eff})
\begin{equation}
\label{Delta-cont}
\Delta_{ab} \approx \alpha \mu_D
={\mathcal O}(\sqrt{n a_{\rm B,0}^3})~,
\end{equation}
in the wave equation (\ref{wave-eq}). 
For the shift of the bound state energy levels follows \cite{Ebeling:1986,Ebeling:1989}
\begin{equation}
\label{Delta-En}
\Delta E_{\rm nl} \approx -\frac{\alpha}{2}\mu_D^2 \langle r \rangle_{\rm nl}
= {\mathcal O}({n a_{\rm B,0}^3})~,
\end{equation}
where $a_{\rm B,0}=1/(\alpha~m)$ is the Bohr radius.

The Debye mass  $\mu_D$, equivalent to the inverse of the Debye radius
$r_{\rm D}$ characterizing the effective range of the interaction, depends on
the square root of the density $n(T)$ of charge carriers.
It is this different response of bound states and scattering continuum to
an increase of density and temperature in the medium which leads to the
Mott effect (see, e.g., Refs. in \cite{Mott:1968} and \cite{Redmer:1997})
for electrons in an insulator:
bound states of the Debye potential can only exist when the Debye radius is
larger than
%\begin{equation}
%\label{Mott-cond}
$r_{\rm D,Mott}=0.84 ~a_{\rm B,0}~~$  \cite{Rogers:1970}.
%\end{equation}
This entails that above a certain density even the ground state electrons
become unbound and form a conduction band, resulting in an insulator-metal
transition also called Mott-transition.
Further details concerning this example can be found in 
Ref.~\cite{Kraeft:1990}.

In complete analogy to this electronic Mott effect it is expected that
in hadronic matter under compression the hadrons as bound states of quarks
undergo a Mott transition which results in
a phase transition from the color insulating phase of hadronic matter to a
color conducting or even color superconducting phase of deconfined quark
matter. This applies to light hadrons as well as to heavy quarkonia, whereby
due to the different scales of Bohr radii the Mott dissociation of heavy
quarkonia occurs at higher densities than for light hadrons.

In most approaches the quark self energy effects
are neglected and one is left with the only medium effect due to a statically
screened potential. This has the consequence that in such a picture the
continuum edge of the scattering states remains unshifted and due to the lack
of compensation the effective interaction leads to a strong medium dependence
of the bound state energies (masses). For the electron-hole plasma in highly
excited semiconductors it could be shown experimentally, however, that the
compensation picture is correct and the bound state energies remain almost
unshifted \cite{Fehrenbach:1981}.

One may of course absorb the self-energy effects into a redefinition of the
effective interaction, by adding a homogeneous mean-field contribution.
This is equivalent to the use of the Ecker-Weitzel potential \cite{Ecker:1956}
\begin{equation}
\label{E-W}
V_{\rm Ecker-Weitzel}(r) = -\frac{\alpha}{r}{\rm e}^{-\mu_D r} - \alpha \mu_D~.
\end{equation}
It is interesting to note that recent investigations of the screening problem
in the context of Debye-H\"uckel theory \cite{Dixit:1989vq}
%for confining potentials
and $Q\bar Q$ correlators \cite{Brambilla:2008cx,Beraudo:2007ky}
have obtained this continuum shift ($-\alpha \mu_D$) as a homogeneous
background field contribution.
According to the above lesson from plasma physics, however, this contribution
should be attributed to the  self-energy of the constituents rather than
to the interaction kernel, since it determines the shift of the continuum edge.

For the development of a comprehensive approach to heavy quarkonia in
hadronizing hot, dense QCD matter another insight from plasma physics
may be of relevance and will be discussed next:
the effect of strong correlations (bound states) in the medium.
To this end, the bound states will be treated like a new species occuring in 
the system. Accordingly, additional diagrams have to be taken into account
which stem from a cluster expansion of the interaction kernel $K_{ab}$ and the 
corresponding self-energy $\Sigma_a$, see Figs. 
\ref{fig:cluster-ex}-\ref{fig:cluster-screen}. 
This leads in the plasma Hamiltonian 
$H^{\rm pl}$ to a generalization of the self-energy contributions 
(cluster-Hartree-Fock approximation), the distribution functions 
in the Pauli-blocking factors and the dynamical screening (cluster-RPA). 
The van-der-Waals interaction in Eq.~(\ref{plasmaH}) appears naturally as a 
contribution in the cluster expansion, describing polarization effects due to 
bound states in the medium.

\section{Cluster expansion for quarkonia in correlated medium}

In the vicinity of the plasma phase transition, correlations play an important
role and their proper accounting requires rather sophisticated theoretical
methods such as cluster expansion techniques.
For the problem of charmonium in dense hadronic matter at the deconfinement
transition, i.e. in the strong coupling case, we suggest a systematic Born
series expansion of collisions with free and bound states in the surrounding
matter so that all terms linear in the density of free particles and bound
states are taken into account.

We describe the cluster expansion here in terms of its diagrammatic
expressions for the interaction kernel and the corresponding self-energy.
\begin{figure}[!t]
%\begin{tabular}{cc}
%\parbox{\textwidth}{
\includegraphics[width=0.95\textwidth,angle=0]{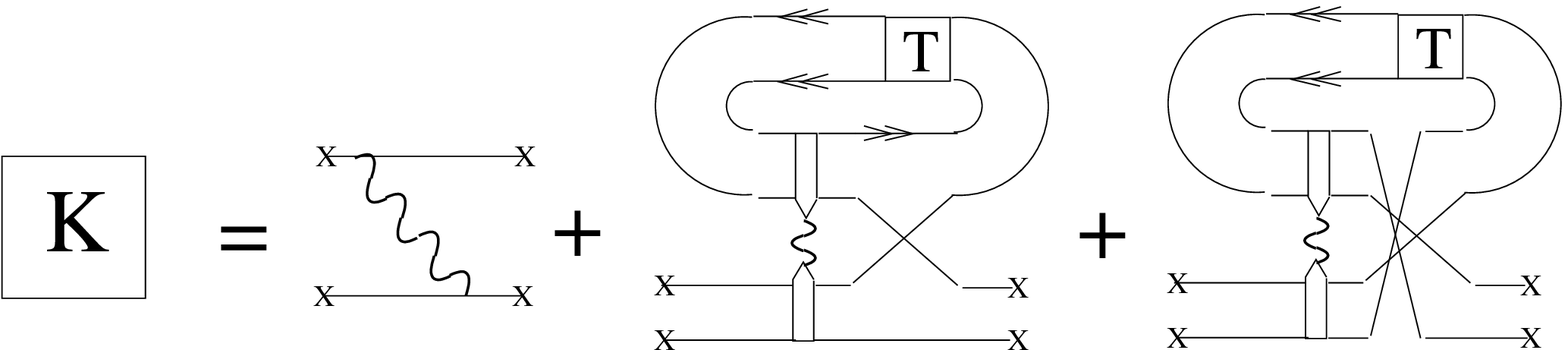}
\\[3mm]
%}
%\parbox{0.6\textwidth}{
\includegraphics[width=0.67\textwidth,angle=0]{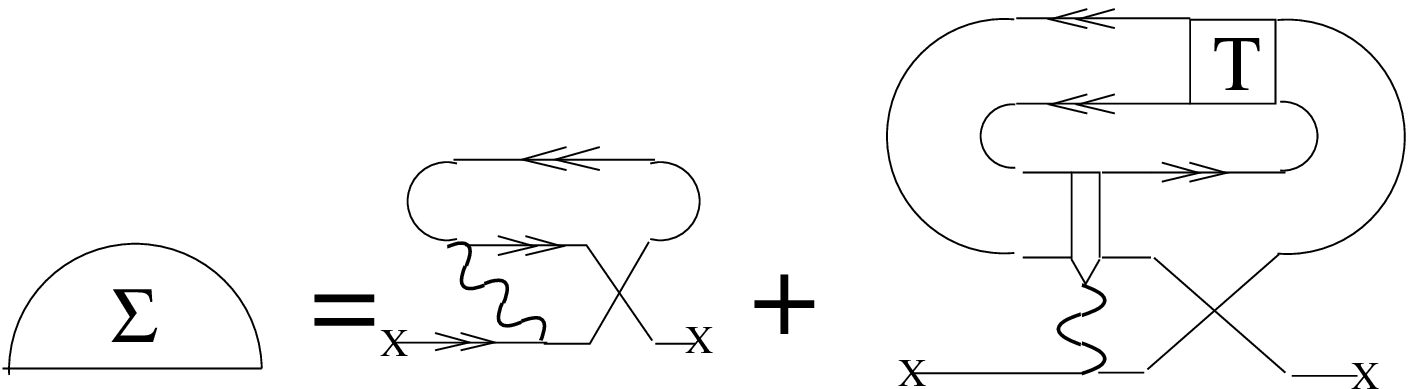}
%}\hfill
%&
%\parbox{0.3\textwidth}{
~~;~~\includegraphics[width=0.25\textwidth,angle=0]{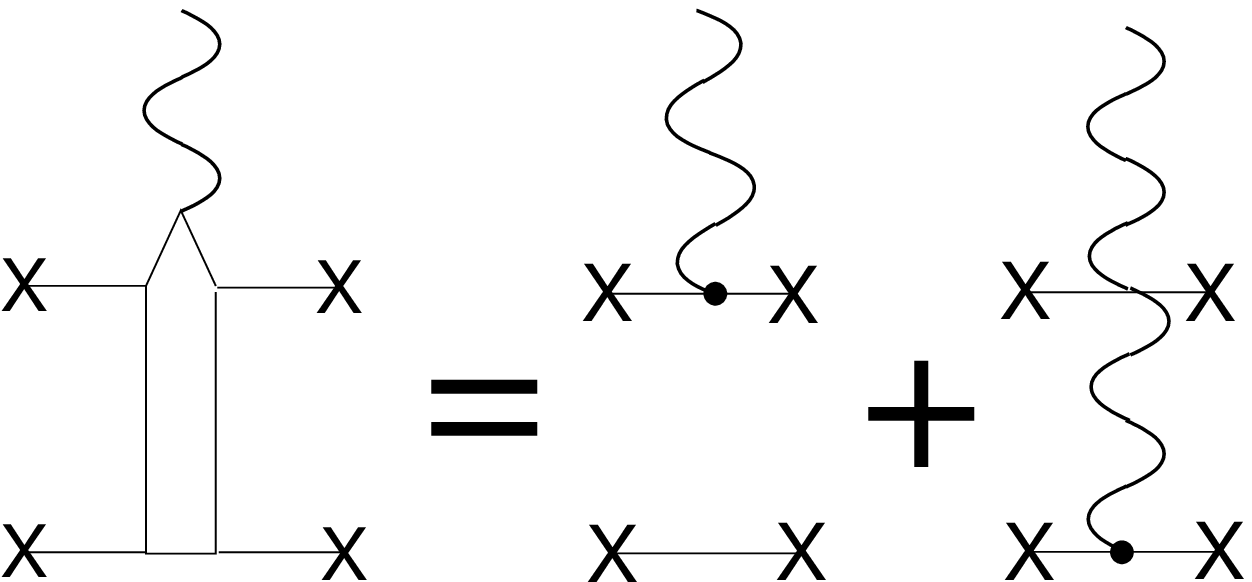}
%}
%\end{tabular}
\caption{Cluster expansion for interaction kernel for the two-particle problem 
in a strongly correlated medium (upper equation) and the corresponding
self-energy (lower left equation) with a dipole ansatz for the vertex 
(lower right equation).
\label{fig:cluster-ex} }
\end{figure}
\begin{figure}[!t]
%\begin{tabular}{cc}
%\parbox{14cm}{
\includegraphics[width=0.95\textwidth,angle=0]{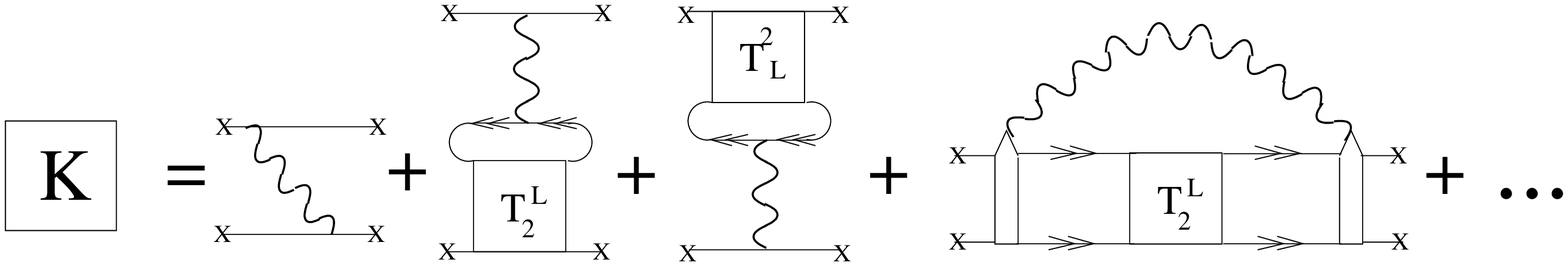}
\\[2mm]
%}
%\parbox{14cm}{
\includegraphics[width=0.95\textwidth,angle=0]{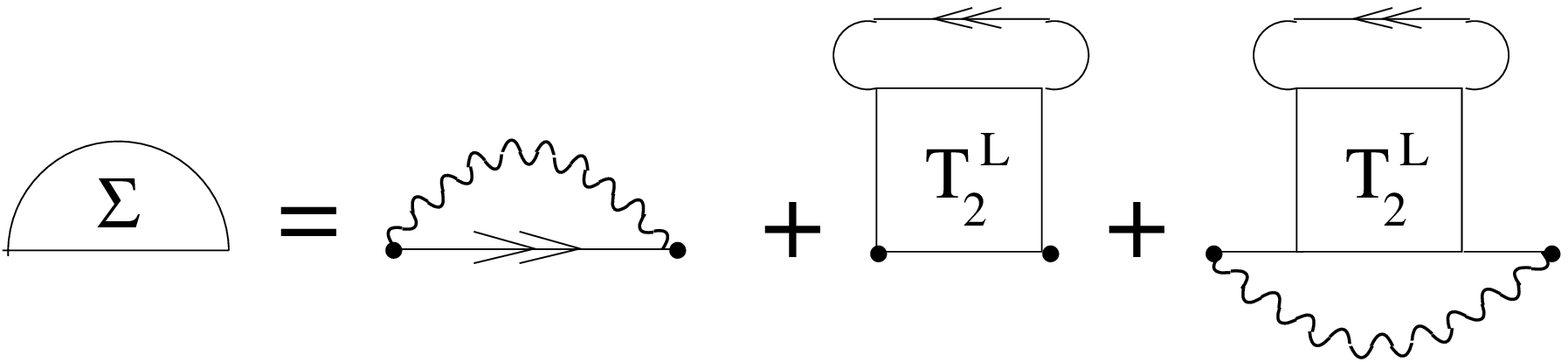}
%}
%\end{tabular}
\caption{Alternative way of drawing the diagrams for the cluster expansion 
of the interaction kernel and the corresponding self-energy of 
Fig.~\ref{fig:cluster-ex} in a form familiar in plasma physics and nuclear
physics. 
\label{fig:cluster-ex2} }
\end{figure}
\begin{figure}[!t]
%\begin{tabular}{cc}
%\parbox{6cm}{
\includegraphics[width=0.35\textwidth,angle=0]{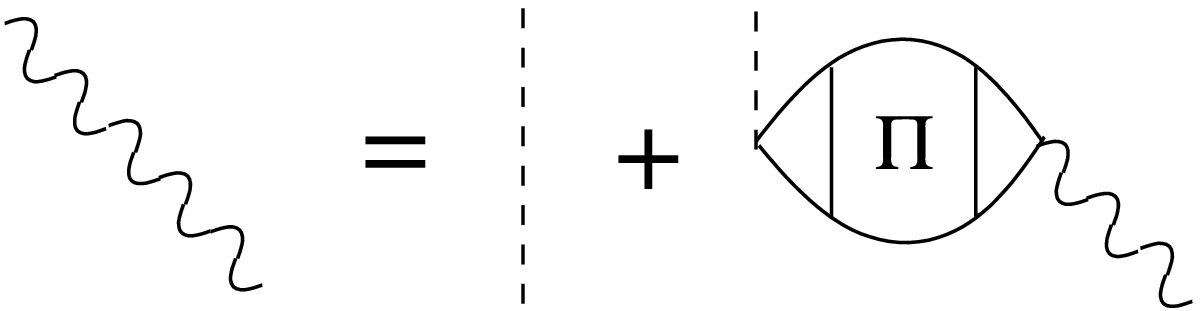}
%}\hfill
%&
%\parbox{9cm}{
~~;~~\includegraphics[width=0.6\textwidth,angle=0]{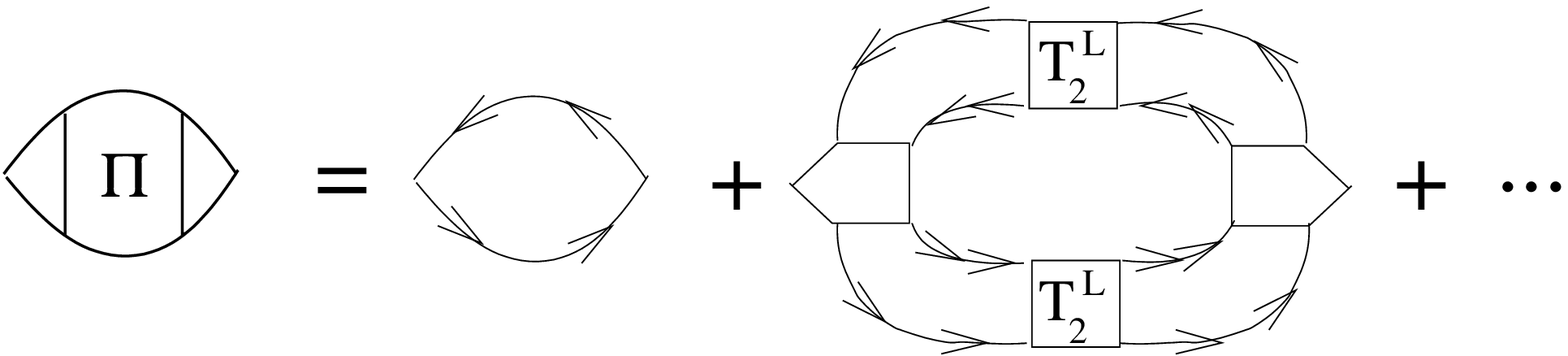}
%}
%\end{tabular}
\caption{Left panel: The dynamically screened interaction potential 
$V_{ab}^S(\omega, q)$ (wavy line), determined by the bare potential 
(dashed line) and the polarization function $\Pi_{ab}(\omega, q)$.
Right panel: Cluster expansion for the generalized RPA, when besides free 
particles (RPA) also two-particle states (cluster-RPA) contribute to the 
polarizability of the medium, see Ref.~\cite{Ropke:1979}.
\label{fig:cluster-screen}}
\end{figure}
The $1^{st}$ Born approximation diagrams of this expansion are given
in Fig. \ref{fig:cluster-ex}, see also the monograph  \cite{Ebeling:1986}.
The wavy line denotes the dynamically screened interaction $V_{ab}^S$,
which in a strongly correlated plasma receives contributions from the 
polarization of the medium beyond the RPA, denoted as generalized (cluster-)
RPA in Fig. \ref{fig:cluster-screen}, see Ref.~\cite{Ropke:1979}.
Bound and scattering states are described consistently in the two-particle 
T-matrices. For a generalization to higher n-particle correlations, see
\cite{Ropke:1983,Ropke:1984} and the monograph \cite{Ebeling:1986}.
The diagrams containing T-matrices do not contribute to the charmonium 
spectrum as long as the densities of the charmed quarks and of charmed hadrons 
in the medium are negligible. 
This is the situation expected for the rather low-energy CBM experiment. 
For the discussion of charmonium production at RHIC and at LHC the inclusion 
of these terms can be invoked.

\begin{figure}[!t]
%\begin{tabular}{cc}
%\parbox{7cm}{
\includegraphics[width=0.47\textwidth,height=0.25\textwidth,angle=0]{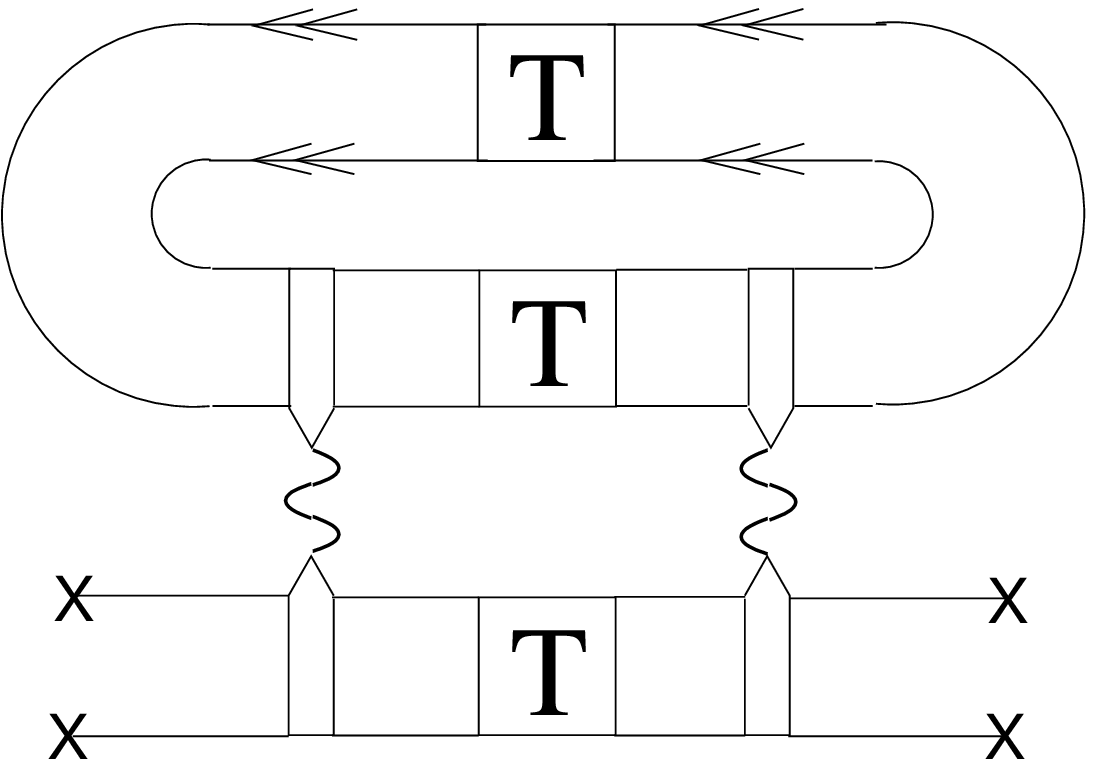}
%}\hfill
%&
%\parbox{7cm}{
~~;~~\includegraphics[width=0.47\textwidth,height=0.25\textwidth,angle=0]
{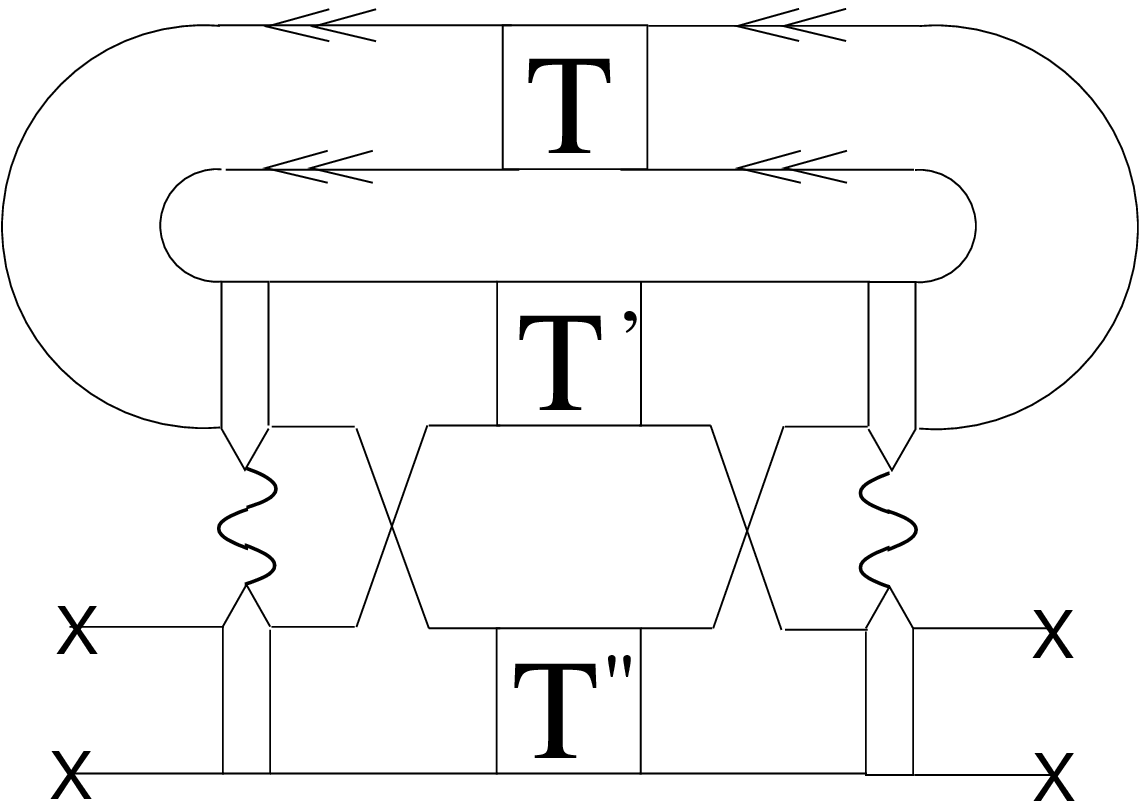}
%}
%\end{tabular}
\caption{Contributions to the dynamical self-energy of a two-particle
system in a correlated medium at  $2^{nd}$ Born order.
Left panel: impact by two-particle states without constituent exchange
(van-der-Waals or dipole-dipole interaction).
Right panel: constituent-rearrangement collisions (string-flip process),
from Ref.~\cite{Blaschke:2006ct}, see also \cite{Satz:2006uh}.
\label{fig:cluster-cluster}
}
\end{figure}

At the $2^{nd}$ Born order, we distinguish two classes of collisions
with light clusters (hadrons) that can give rise to spectral broadening of the
charmonia.
The first class concerns hadron impact without quark rearrangement inducing 
transitions to excited states, shown in the left panel of  
Fig. \ref{fig:cluster-cluster}.
These processes have been considered for charmonium-hadron interactions within 
the operator product expansion techniques following Peskin and Bhanot 
\cite{Bhanot:1979vb,Peskin:1979va}, see \cite{Kharzeev:1994pz,Arleo:2001mp}.
The result is a deformation of the charmonium spectrum under conservation of 
the spectral weight integrated over all charmonia states.
In the second class are quark rearrangement (string-flip) processes,
as indicated in the right panel of Fig. \ref{fig:cluster-cluster}.
They induce transitions to open charm hadrons responsible for charmonium 
dissociation in hadronic matter, cf. Sect. \ref{sec:diss-hg}.

\paragraph{Example 2: String-flip model of charmonium dissociation.}

\begin{figure}[!t]
%\begin{tabular}{cc}
\parbox{0.3\textwidth}{
\includegraphics[width=0.22\textwidth,angle=0]{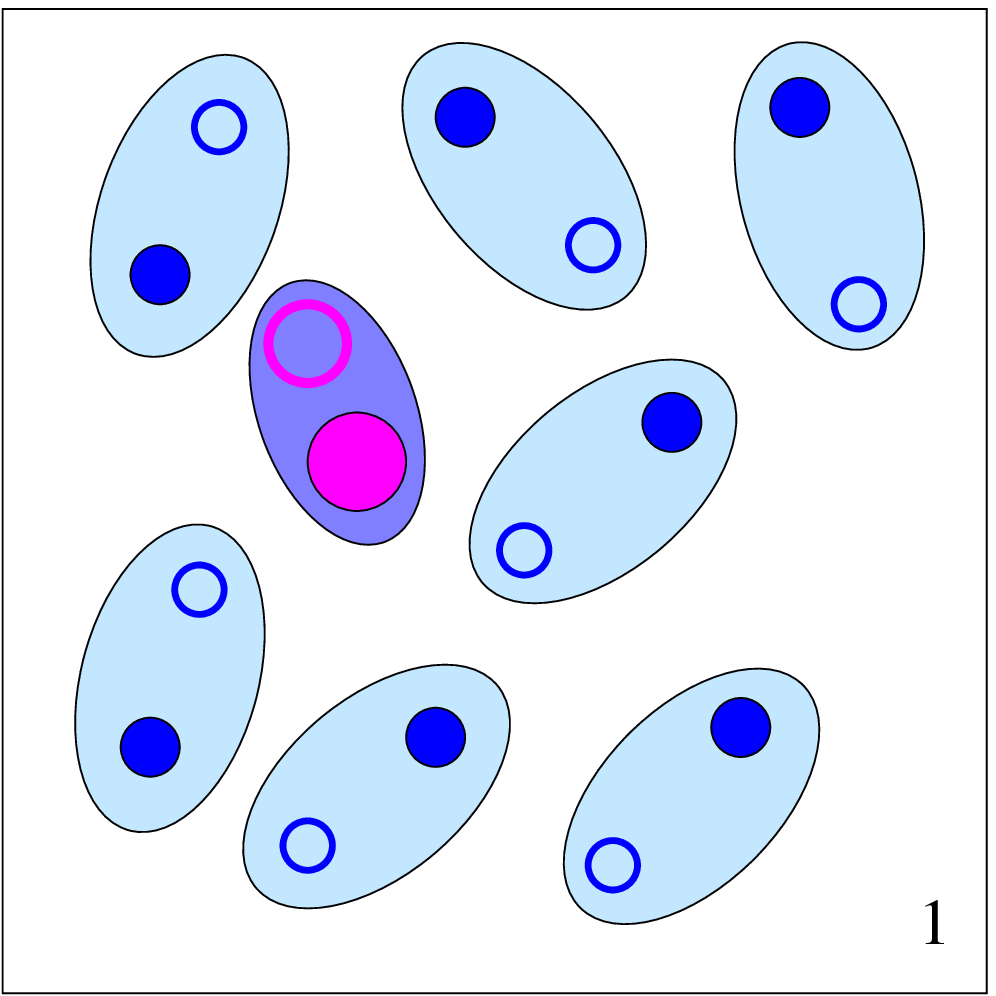}
\includegraphics[width=0.22\textwidth,angle=0]{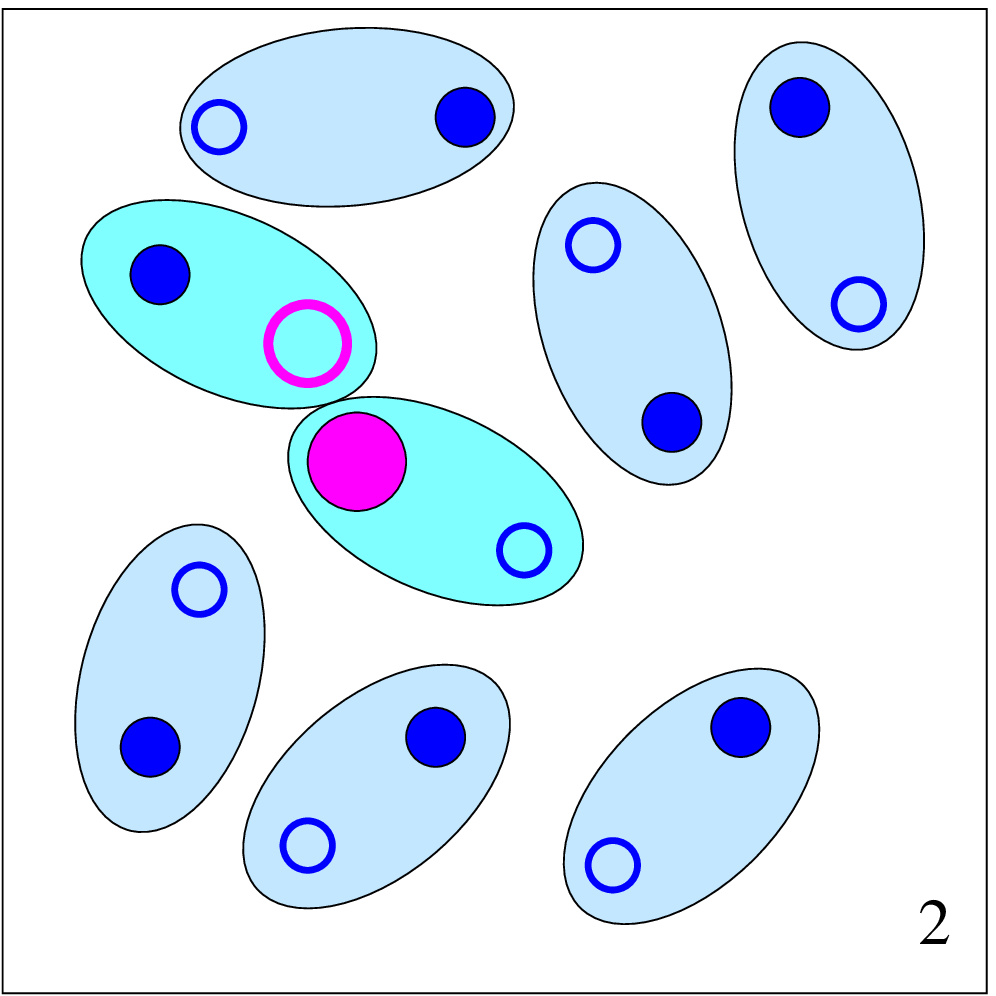}
}%\hfill
%&
\parbox{0.65\textwidth}{
\includegraphics[width=0.7\textwidth,angle=0]{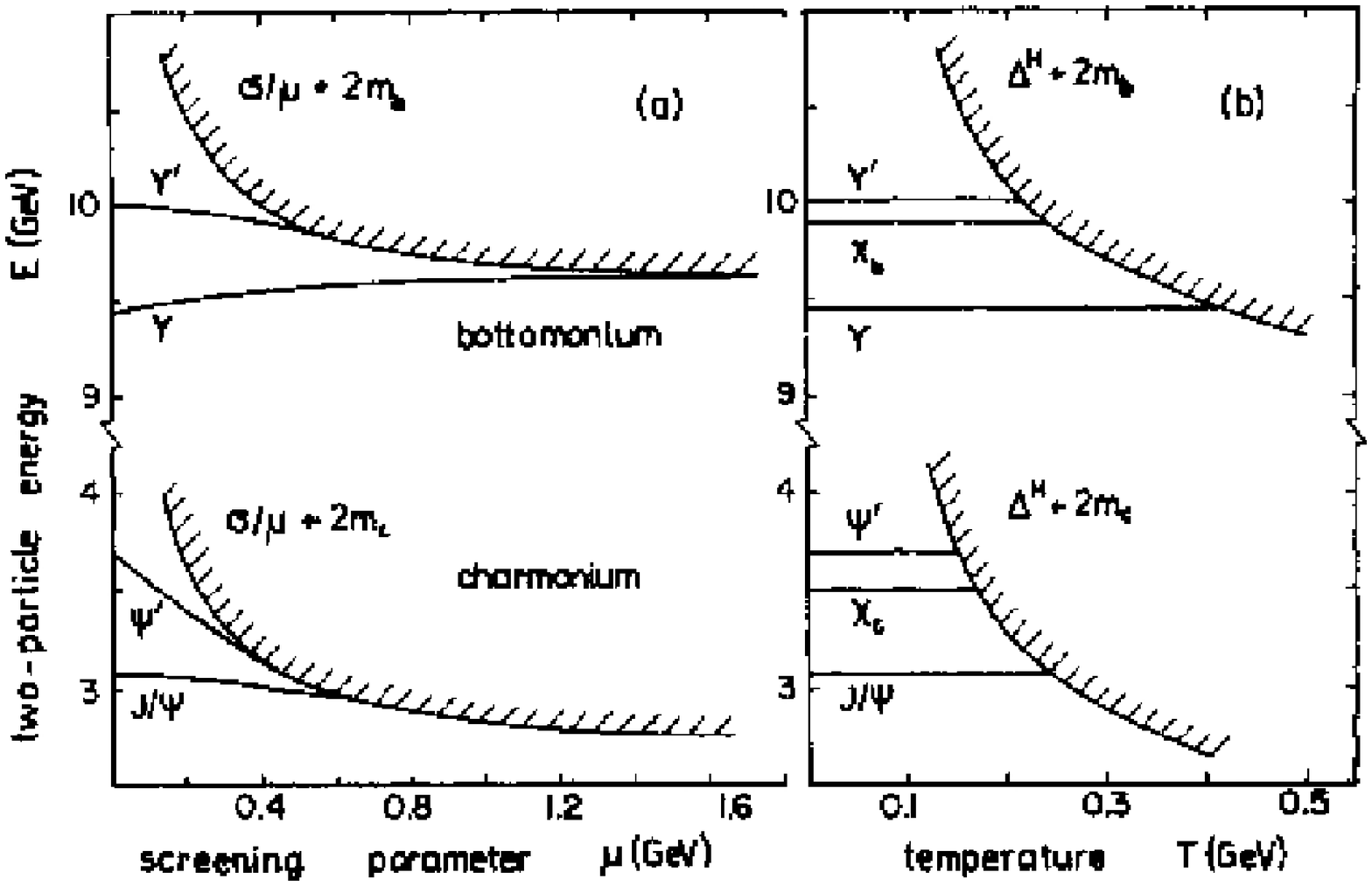}
}
%\end{tabular}
\caption{Left panel: String-flip process in a dense quark-antiquark system, 
see also \cite{Ropke:1986qs,Satz:2006uh}.
Right panel:  Two-particle energies of charmonia and bottomonia states
in a statically screened potential (a) and in the string-flip model (b), from 
Ref. \cite{Ropke:1988zz}. 
\label{fig:string-flip}
}
\end{figure}

Here we give a second example for the use of insights from plasma physics 
by discussing charmonium dissociation within the string-flip model of quark 
matter 
\cite{Horowitz:1985tx,Ropke:1986qs,Miyazawa:1979vx,Blaschke:1984yj}.
In this model the string-type color interactions between quarks get saturated
within the sphere of nearest neighbors so that in a dense system of 
overlapping quark-antiquark pairs frequent string-flip processes take place
in order to assure the system is at any time in its minimal energy 
configuration, see the left panel of Fig.~\ref{fig:string-flip}. 

When  considering a heavy quark-antiquark pair in dense matter with negligible 
heavy-flavor fraction, the Pauli blocking and exchange self-energy 
contributions are negligible, but the strong correlations with light quarks of 
complementary color within the nearest neighbor sphere will result in a 
meanfield selfenergy shift (Hartree shift $\Delta^H$) for all quarks 
\cite{Ropke:1988bx}. 
This determines the shift of
the continuum edge, see the graph (b) in the right panel of 
Fig.~\ref{fig:string-flip}. Because of the compensation in the Bethe-Salpeter
kernel between the effects of the screening of the interaction and the self 
energy shifts calculated with it (see discussion above), it is suggested that 
to lowest order the bound state energies remain unshifted when increasing the 
temperature and/or density of the medium.
In contrast to the first example of Debye screening of long-range Coulombic 
interactions, the screening mechanism in the string flip model is color 
saturation within nearest neighbors, applicable for strong, short-range 
interactions as appropriate for the case of the sQGP at RHIC or dense systems
at FAIR CBM.
The resulting two-particle energy spectrum for charmonium and bottomonium 
states is shown in the right panel of Fig.~\ref{fig:string-flip}, where the 
static screening picture (graph (a) as a function of the screening parameter
$\mu=\mu_D(T)$ in the screened Cornell potential \cite{Karsch:1987pv} is 
compared to the string-flip picture (graph (b) as a function of the 
temperature $T$), from Ref.~\cite{Ropke:1988bx,Ropke:1988zz}. 
From this Figure one can read-off the in-medium lowering of the  
dissociation threshold $k_0^{\rm diss}$, which is the energy difference 
between the considered bound state level and the continuum edge shown as the 
border of the hatched region. 
This lowering of $k_0^{\rm diss}$ with increasing density and/or temperature
leads to a strong increase in the quarkonium breakup cross sections by thermal 
impact \cite{Ropke:1988zz}
and to the bound state dissociation, even before the binding energies vanish 
at the critical Mott densities and temperatures for the corresponding states.

\section{Charmonium dissociation in a resonance gas}
\label{sec:diss-hg}
An interesting phenomenological guideline for the present discussion 
is provided by Ref.~\cite{Prorok:2000kv} where the authors show 
that universal $J/\psi$-hadron breakup cross sections with correct
kinematic thresholds but otherwise constant at 3 mb for meson and 5 mb 
for baryon impact are sufficient to explain the observed anomalous 
$J/\psi$ suppression pattern of the NA50 experiment within a multi-component
hadron gas. For a recent update, see \cite{Prorok:2008zm}.
The question arises: How do these assumptions relate to microscopic
calculations of charmonium dissociation reactions in hadronic matter? 
Those can generally be divided into two categories, based on 
hadronic or quark degrees of freedom.
We will primarily review and compare the studies of processes in 
a mesonic medium (predominantly composed of pions and rho mesons) 
within both approaches, including a discussion of in-medium effects. 

Historically, a first calculation of the quark-exchange
reaction, $J/\psi + \pi \to D + \bar{D}^* + c.c.$, was
performed in a nonrelativistic quark model~\cite{Martins:1994hd}
based on applications of the diagrammatic technique developed by
Barnes and Swanson~\cite{Barnes:1991em}
for meson-meson scattering, see also Ref.~\cite{Blaschke:1992qa}.
This calculation showed a strong energy dependence of the cross section
with a peak value of about 6~mb at threshold and a fast decrease due
to a momentum mismatch in the overlap integrals between the meson wave 
functions.
The thermally averaged cross section, which is the relevant quantity for 
estimating the $J/\psi$ dissociation rate, was below 1~mb, roughly in
accordance with phenomenological expectations based on the 
observed suppression in heavy-ion reactions at the SPS.
\begin{figure}[!t]
\begin{tabular}{cc}
\includegraphics[width=0.47\textwidth,height=0.3\textwidth]{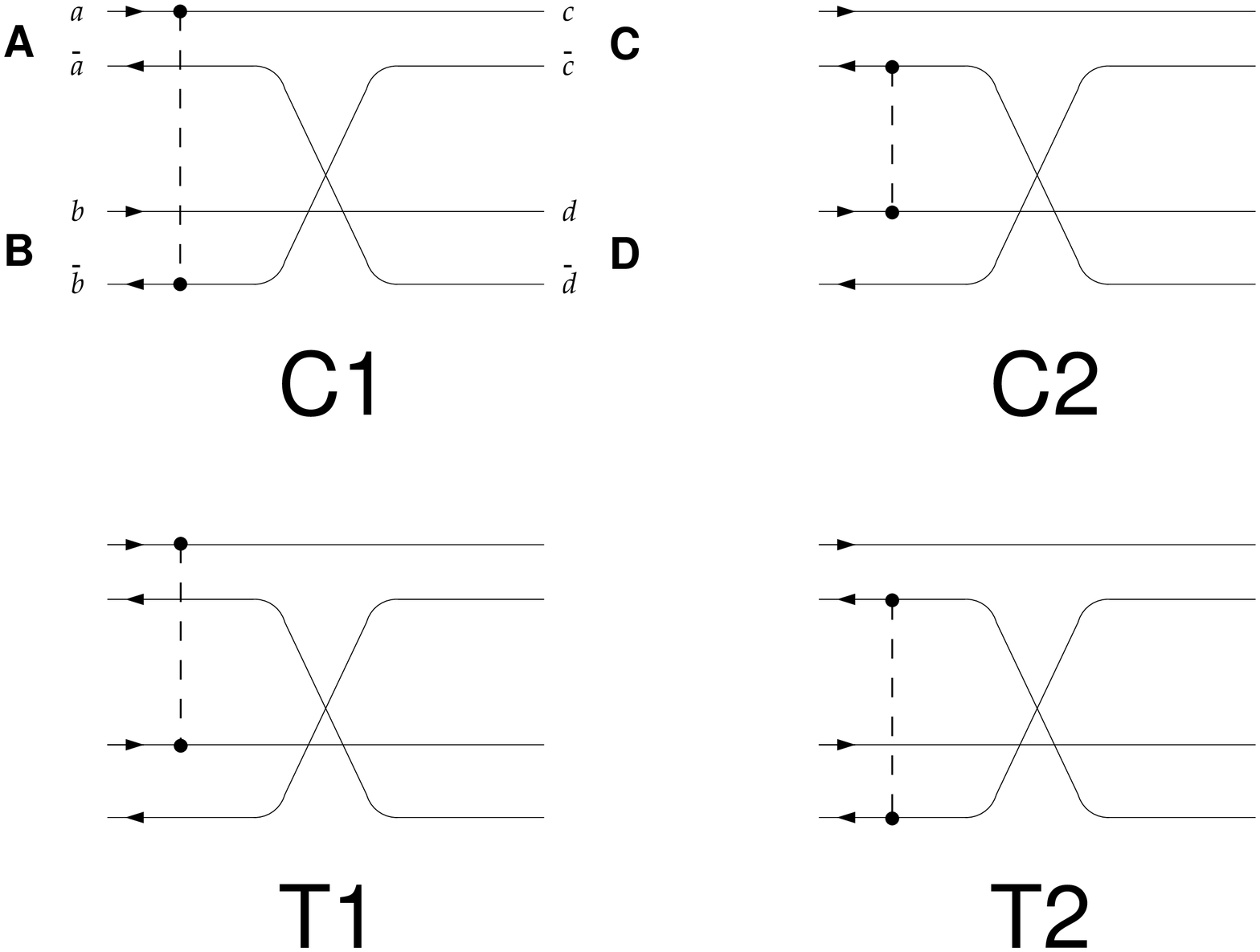}&
\includegraphics[width=0.47\textwidth,height=0.3\textwidth]{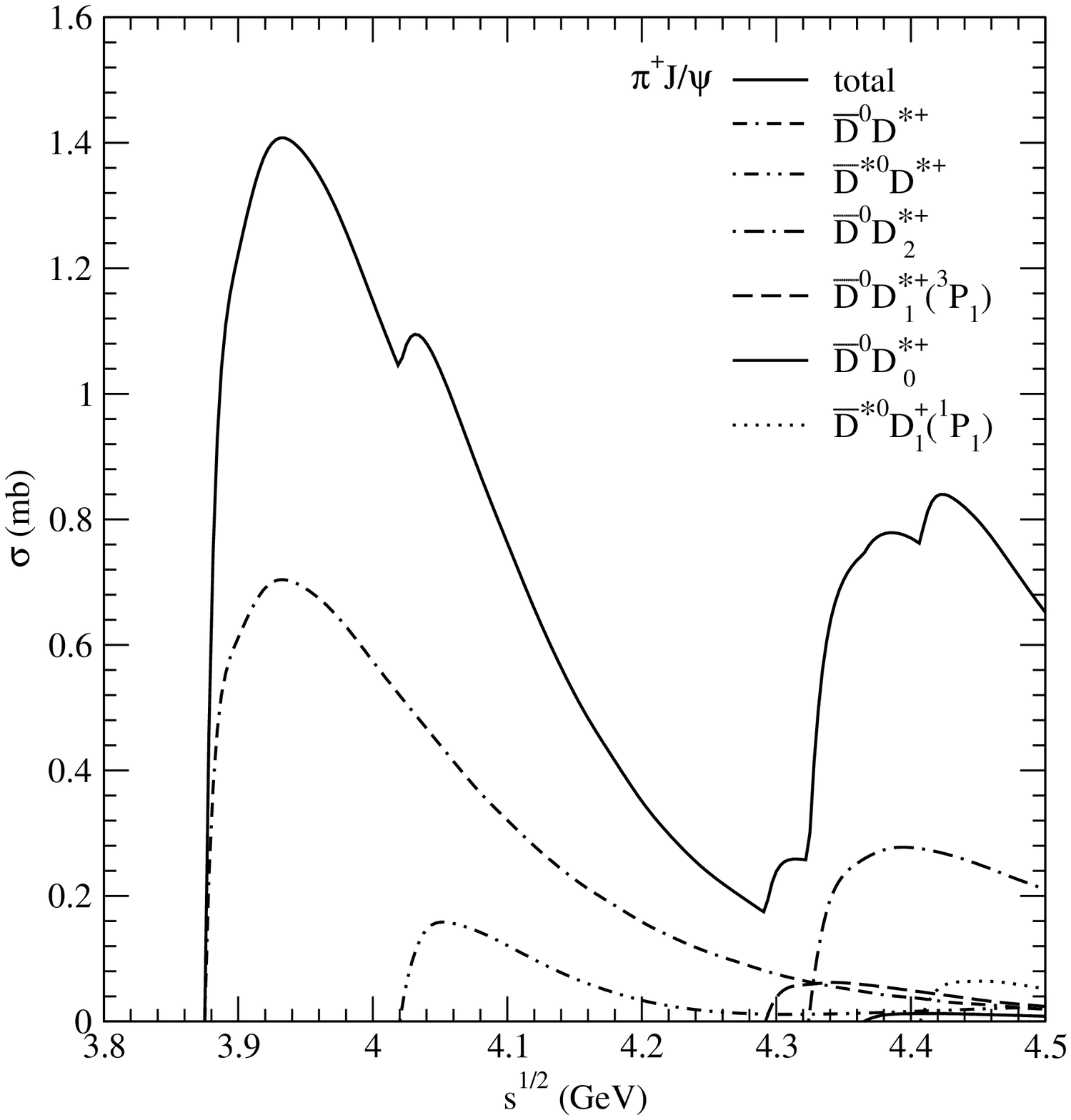}
\end{tabular}
\caption{Born diagrams for quark-exchange processes (denoted as C1, C2 for 
``capture'' and T1, T2 for ``transfer'' diagrams) contributing to
heavy quarkonium dissociation by meson impact (left panels) and the
corresponding cross section for $J/\psi$ breakup by pion impact (right
panel); from Refs.~\cite{Wong:1999zb,Barnes:2003dg}.
Note that in the total cross section also the processes with charge conjugated 
final states are accounted for.
\label{jpsimesondiss}
}
\end{figure}
These calculations were extended to other light mesons and excited charmonia
in Refs.~\cite{Wong:1999zb,Barnes:2003dg},
where also more realistic quark-interaction potentials have been used.
In Fig.~\ref{jpsimesondiss} we show the diagrams for
quark-exchange processes at first Born order, which are classified as
``capture'' (C) and ``transfer'' (T) diagrams depending on whether the quark
interaction can be absorbed into the external meson lines; the
latter are to be understood as a resummation of  
ladder-type quark-antiquark interactions.
There are cancellations among the contributions of the different diagrams
due to the small color dipole of the charmonium state. These cancellations
reduce the peak value of the cross section to about 1~mb, as 
also illustrated in Fig.~\ref{jpsimesondiss}.
An open question in this approach is whether the double nature of the pion
-- a Goldstone boson of the broken chiral symmetry and a strongly bound
quark-antiquark -- has a strong influence on these results. 
In the nonrelativistic approach the pion emerges due to a large hyperfine 
splitting in the Fermi-Breit Hamiltonian (as opposed to, e.g., 
instanton-induced interactions), which is not a robust interpretation.
Another question concerns the applicability of the truncation of the 
transfer diagram contributions at the first Born order.
Ladder-type resummations would lead to $s$- and $t$-channel $D$-meson exchange
processes, which are disregarded in the nonrelativistic quark exchange.
Finally, these models are beset with the ``post-prior'' problem due to
the ambiguity in the ordering of quark exchange and interaction lines.

\begin{figure}[!t]
\parbox{0.5\textwidth}{
\begin{tabular}{cc}
\includegraphics[width=0.25\textwidth,height=0.13\textwidth]{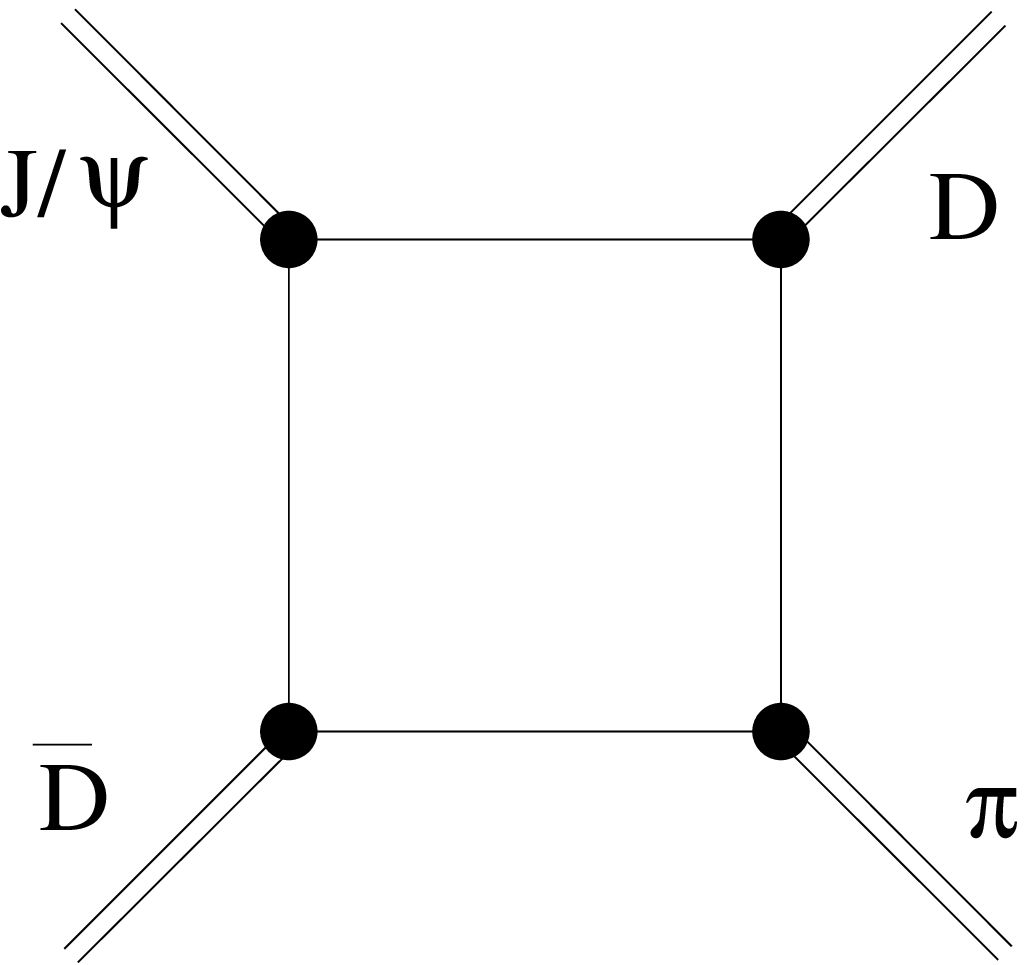}&
\includegraphics[width=0.25\textwidth,height=0.13\textwidth]{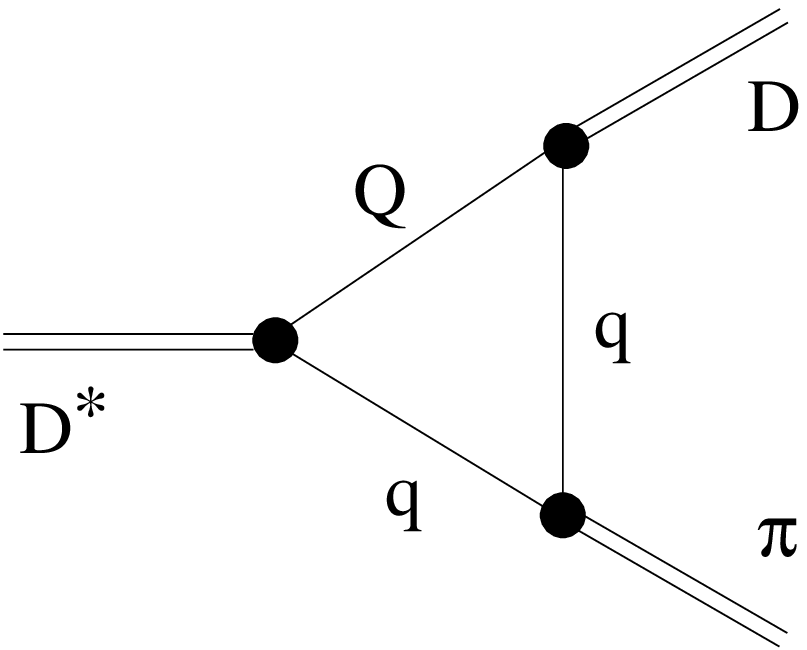}
\end{tabular}
\\[3mm]
\includegraphics[width=0.54\textwidth,height=0.16\textwidth]{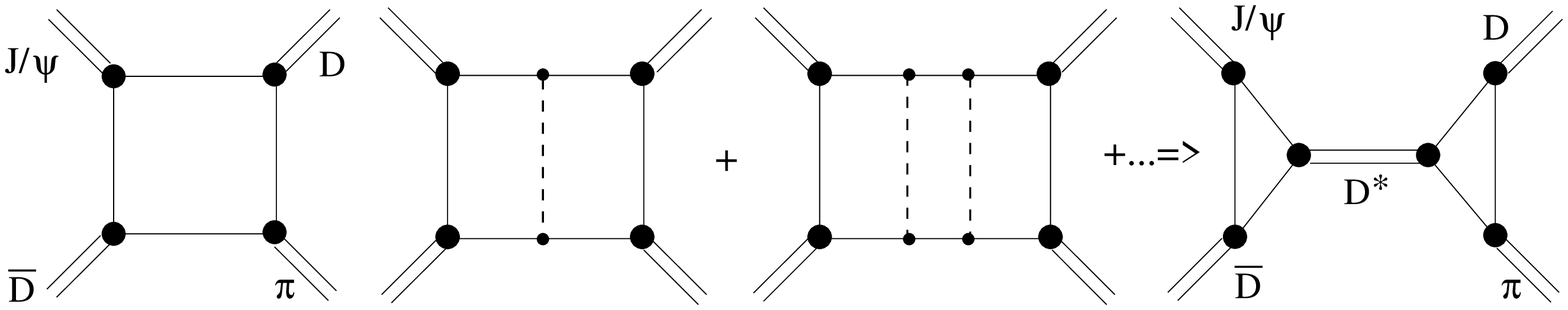}
}%\hfill
\hspace{-0.5cm}
\parbox{0.45\textwidth}{
\includegraphics[width=0.5\textwidth,height=0.4\textwidth]{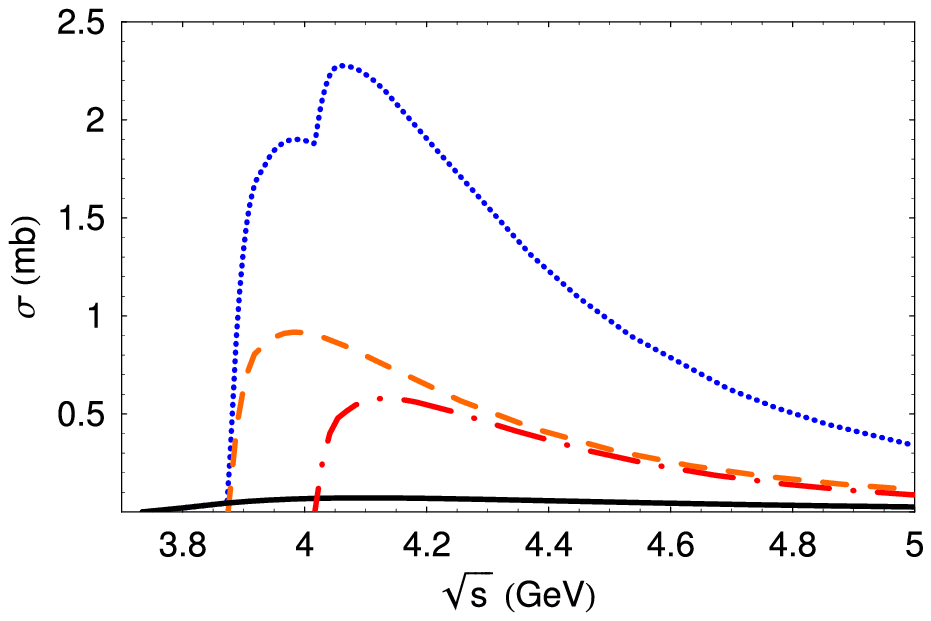}
}
\caption{Left panel: box and triangle diagrams for meson-meson
interaction vertices contributing to $J/\psi$ breakup by meson impact in
the relativistic quark model~\cite{Blaschke:2000zm} (upper part),
and the origin of $D$-meson exchange from ladder-type resummation (lower part).
Right panel: cross section for pion induced $J/\psi$ dissociation into
open-charm mesons (dotted line) composed of subprocesses with different
final-state $D$-meson pairs: $D+\bar{D}$ (solid), $D+\bar{D}^*$ (dashed),
 $D^*+\bar{D}^*$ (dash-dotted); from Ref.~\cite{Ivanov:2003ge}.
The total cross section includes also the subprocess with $\bar{D}+{D}^*$ 
final state for which the cross section is identical to the charge conjugated 
one (dashed). 
\label{jpsipidiss}
}
\end{figure}

These problems can be resolved within relativistic quark models
developed on the basis of Dyson-Schwinger equations~\cite{Roberts:1994dr} 
for applications to the charmonium dissociation 
problem~\cite{Blaschke:2000zm,Ivanov:2003ge,Bourque:2008es}.
In this approach, the meson-meson interactions are represented by quark-loop
diagrams with three (triangle) and four (box) meson legs. The appearance of
meson-exchange processes can be understood as  a ladder resummation
of quark interaction diagrams in $s$- and $t$-channels, see
Fig.~\ref{jpsipidiss}.
The results for the $J/\psi$ dissociation cross section by pion impact
within the relativistic quark model~\cite{Ivanov:2003ge}, shown in
Fig.~\ref{jpsipidiss}, basically confirm those of the nonrelativistic
approaches~\cite{Martins:1994hd,Wong:1999zb,Barnes:2003dg} shown in 
Fig.~\ref{jpsimesondiss}; residual 
differences may be traced back to the treatment of the transfer-type 
diagrams. In the relativistic treatment these diagrams are resummed beyond the
first Born approximation and result in $D$-meson exchange diagrams which are
absent in the nonrelativistic models. This difference may explain small 
differences in cross section results from both approaches.

\begin{figure}[!t]
%\begin{tabular}{cr}
%\includegraphics[width=7.5cm,height=6cm]{diagrams.eps}&
%\includegraphics[width=8.5cm,height=6cm]{sigma-ff.eps}
\parbox{0.5\textwidth}{
\includegraphics[width=0.5\textwidth,height=0.4\textwidth]{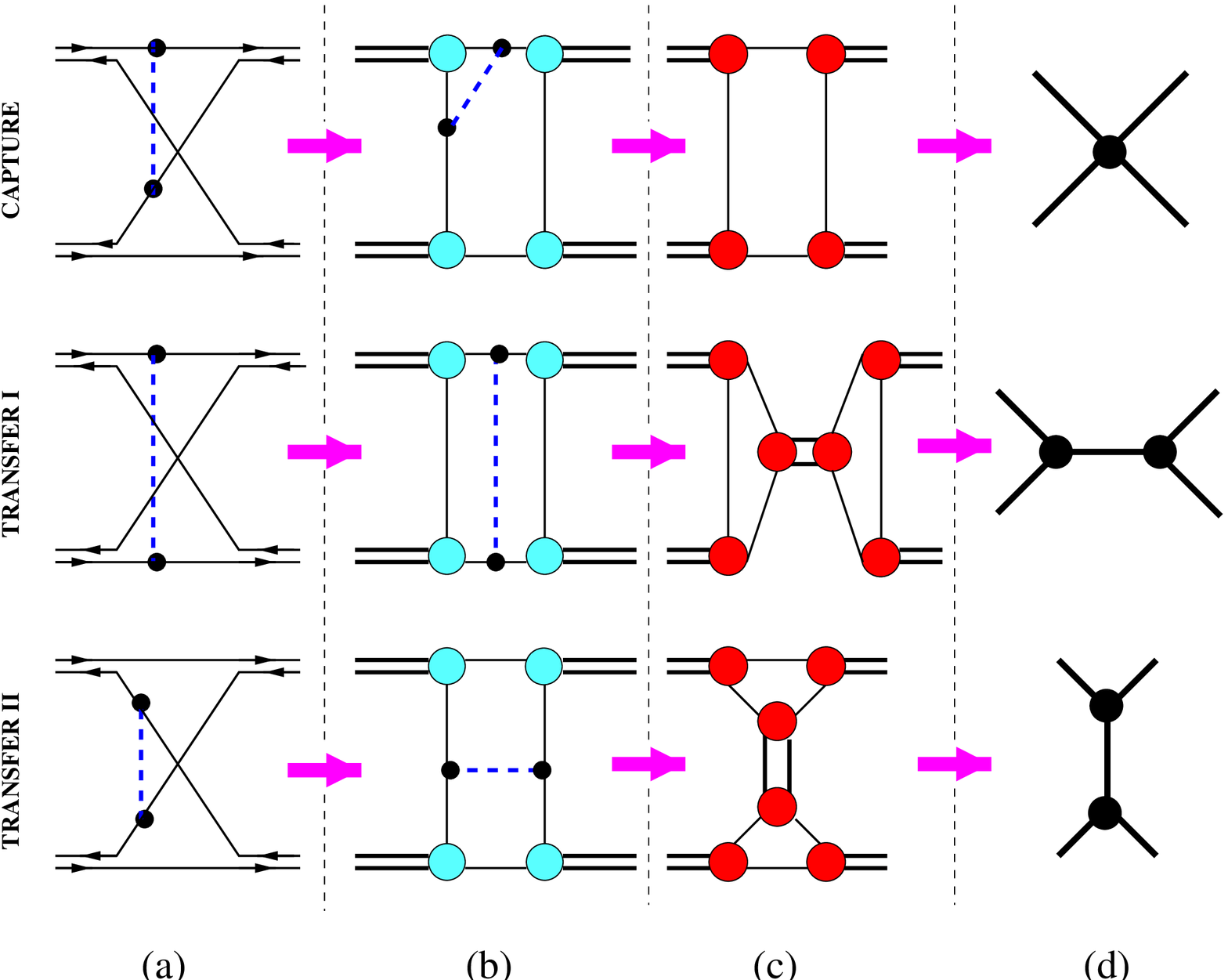}
}\hfill
\parbox{0.45\textwidth}{
\includegraphics[width=0.5\textwidth,height=0.5\textwidth,angle=-90]{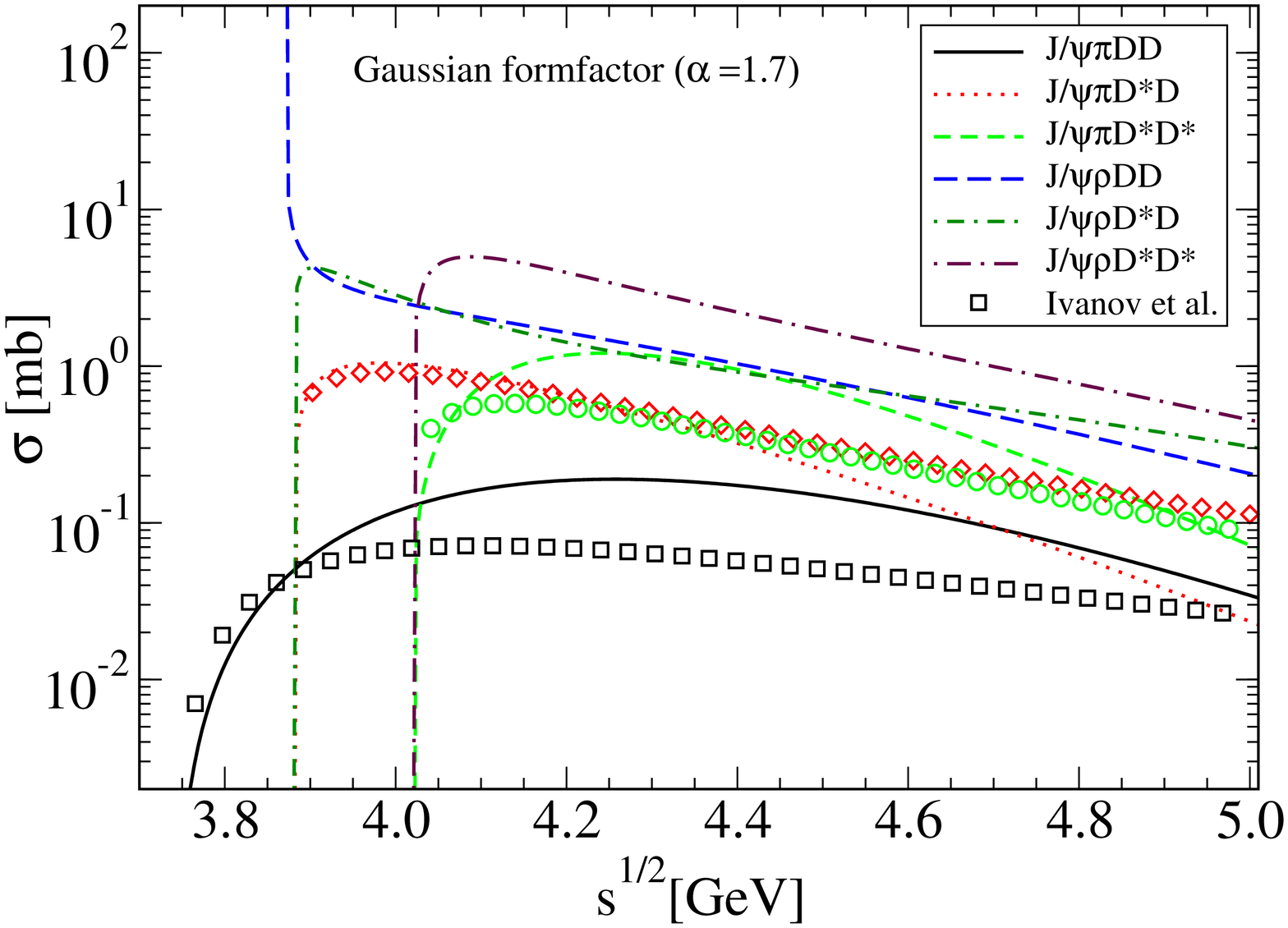}
}
%\end{tabular}
\caption{Feynman diagrams for $J/\psi$ breakup by meson impact
in the chiral Lagrangian approach (left) and the resulting cross sections
for pion- and rho-meson induced processes (right),
from Ref.~\cite{Blaschke:2008mu}.
The left panel illustrates the interrelation of the different approaches to 
charmonium breakup cross sections: The nonrelativistic potential model 
diagrams (a) can be redrawn as quark loop diagrams with Born order insertions 
(b). The latter can be either absorbed into the nonperturbative 
meson-quark-antiquark vertices (c, upper line) or after partial resummation of
all ladder-type diagrams of the Born series redrawn as t-channel and s-channel
meson exchanges (c, lower lines). The locan limit of the relativistic quark 
model diagrams (c) leads to the chiral Lagrangian diagrams (d) with formfactors
originating from the quark loop diagrams for the meson vertices in (c).
\label{jpsidiss-chiLa}
}
\end{figure}

Already in 1998, Matinyan and M\"uller~\cite{Matinyan:1998cb} pioneered an
alternative approach to charmonium absorption by light mesons on the basis of
an effective meson Lagrangian, with a local $U$(4) flavor symmetry strongly 
broken by the pseudoscalar and vector meson mass matrices.
This initial  version of the chiral Lagrangian approach gave rather small
cross sections, $\sigma_{\pi\psi}\approx 0.3$~mb at threshold.
It turned out~\cite{Haglin:1999xs,Lin:1999ad} that triple vector-meson 
couplings and contact terms, not included in Ref.~\cite{Matinyan:1998cb},
result in an increase of the breakup cross section by up two orders of 
magnitude, with a rising energy dependence.
A problem for the chiral Lagrangian approaches is the treatment
of hadrons as pointlike objects, which, as usual for an effective theory, 
becomes unreliable at high momentum transfers (due to quark-exchange
substructure effects). The composite nature of hadrons can be accounted
for in an approximate way by vertex form 
factors~\cite{Lin:1999ad,Haglin:2000ar}, which reduce 
the magnitude, and affect the energy dependence, of the cross sections,
see Fig.~\ref{jpsidiss-chiLa}.
The choice of the cutoff parameters in the form factors has a 
large effect on the charmonium 
breakup cross sections and remains a matter of 
debate~\cite{Oh:2000qr,Ivanov:2001th,Oh:2002vg,Bourque:2004av}.
Progress may be made by calibrating the form factors with
microscopic approaches as, e.g., the nonrelativistic or
the relativistic quark models \cite{Blaschke:2008mu}.
While the calculations with the relativistic quark model are very cumbersome,
the chiral Lagrangian models offer a very effective tool to assess many 
other dissociation processes required for phenomenology, once the
formfactor question is settled.
In particular, these applications include dissociation processes by nucleon
impact~\cite{Sibirtsev:2000aw,Liu:2001ce} and bottomonium 
dissociation~\cite{Lin:2000ke}.

\begin{figure}[!t]
%\parbox{9cm}{
\begin{tabular}{cc}
\includegraphics[width=0.47\textwidth,height=0.3\textwidth]{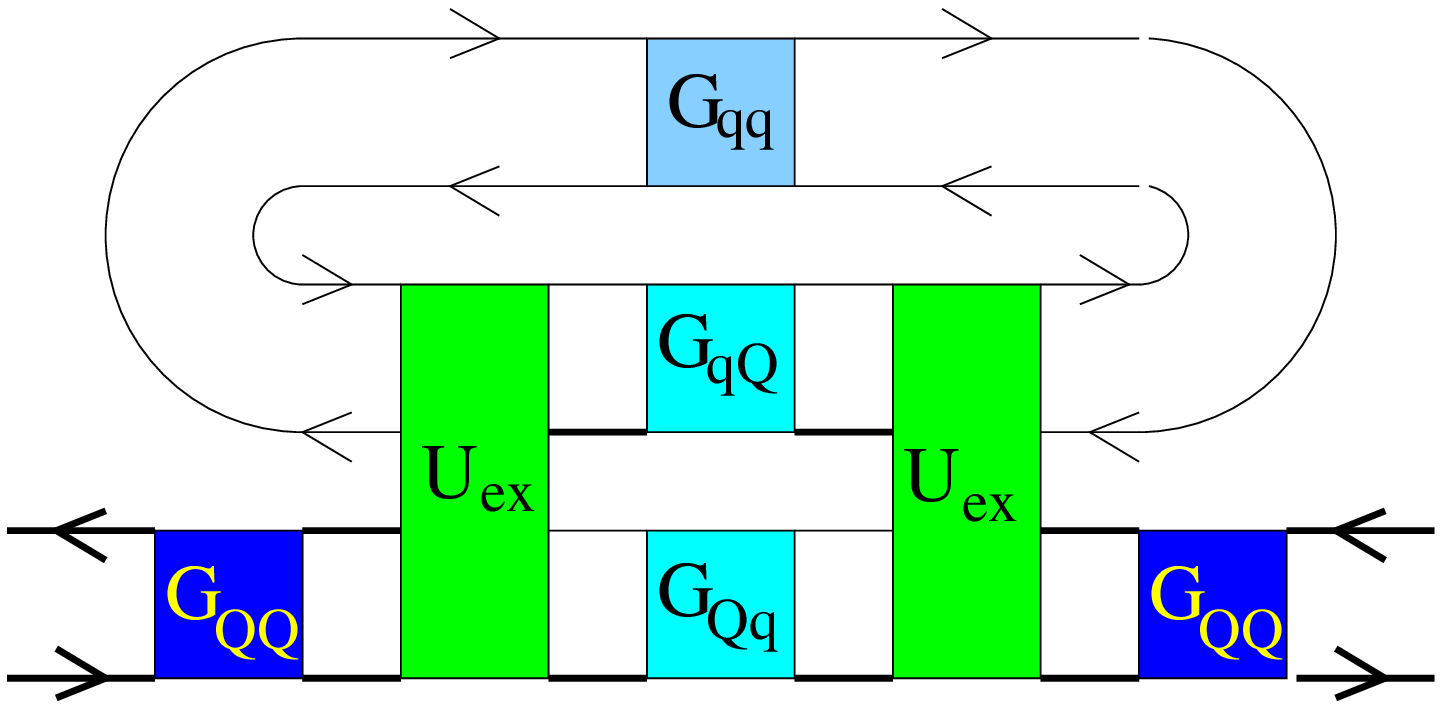}&
%}\hfill
%\parbox{9cm}{
\includegraphics[width=0.47\textwidth,height=0.3\textwidth]{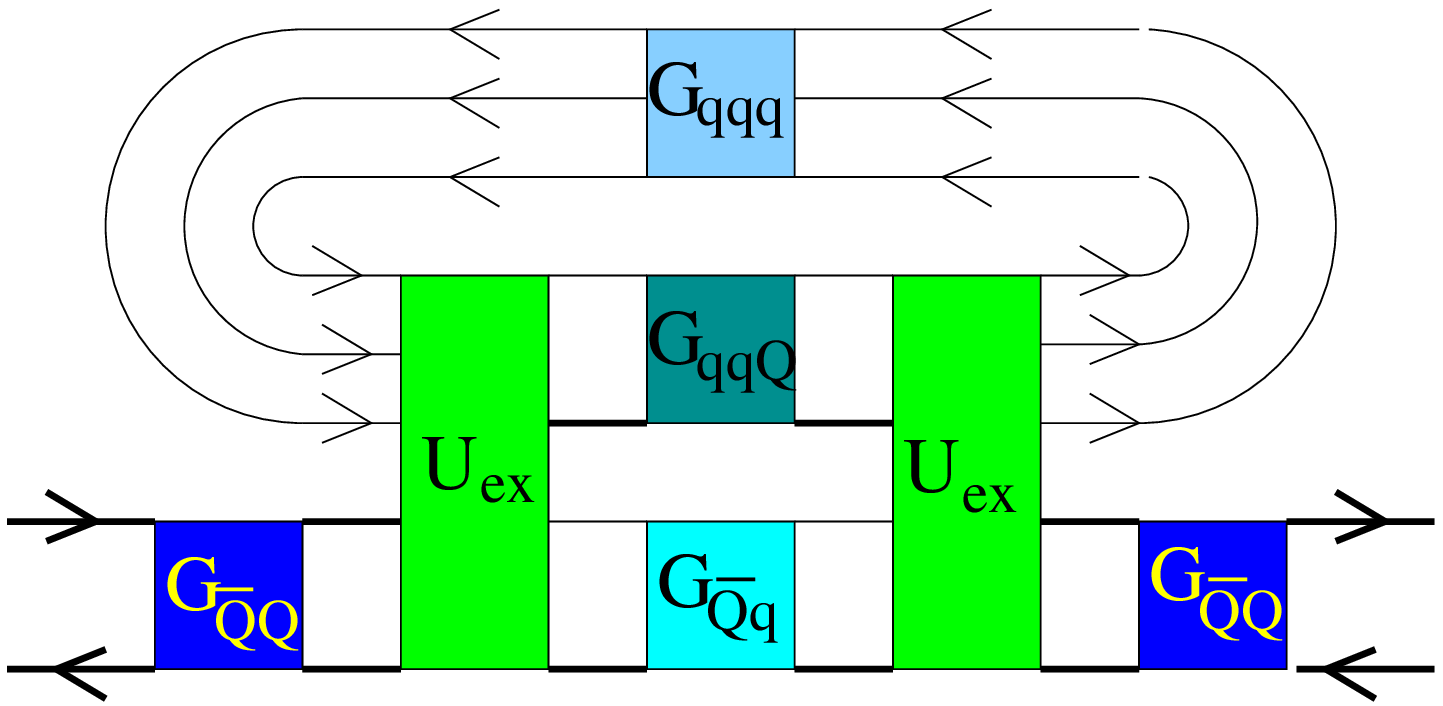}
\end{tabular}
%}
\caption{Left panel: Diagrammatic representation of the quark exchange process
contribution to the heavy quarkonium self-energy in a mesonic 
medium~\cite{Blaschke:2000er}; 
right panel: same as left panel but for quark exchange with baryons in the 
medium~\cite{Blaschke:2004dv}. 
Three different approaches to the interaction vertex $U_{\rm ex}$ are discussed
in the text and shown in Figs.~\ref{jpsimesondiss}-\ref{jpsidiss-chiLa}.
Diagrams of this type appear in the cluster expansion for two-particle 
properties, see Fig.~\ref{fig:cluster-cluster}.
\label{jpsidiss-diagram}
}
\end{figure}

\begin{figure}[!t]
%\parbox{9cm}{
\begin{tabular}{cc}
\includegraphics[width=0.47\textwidth,height=0.35\textwidth]{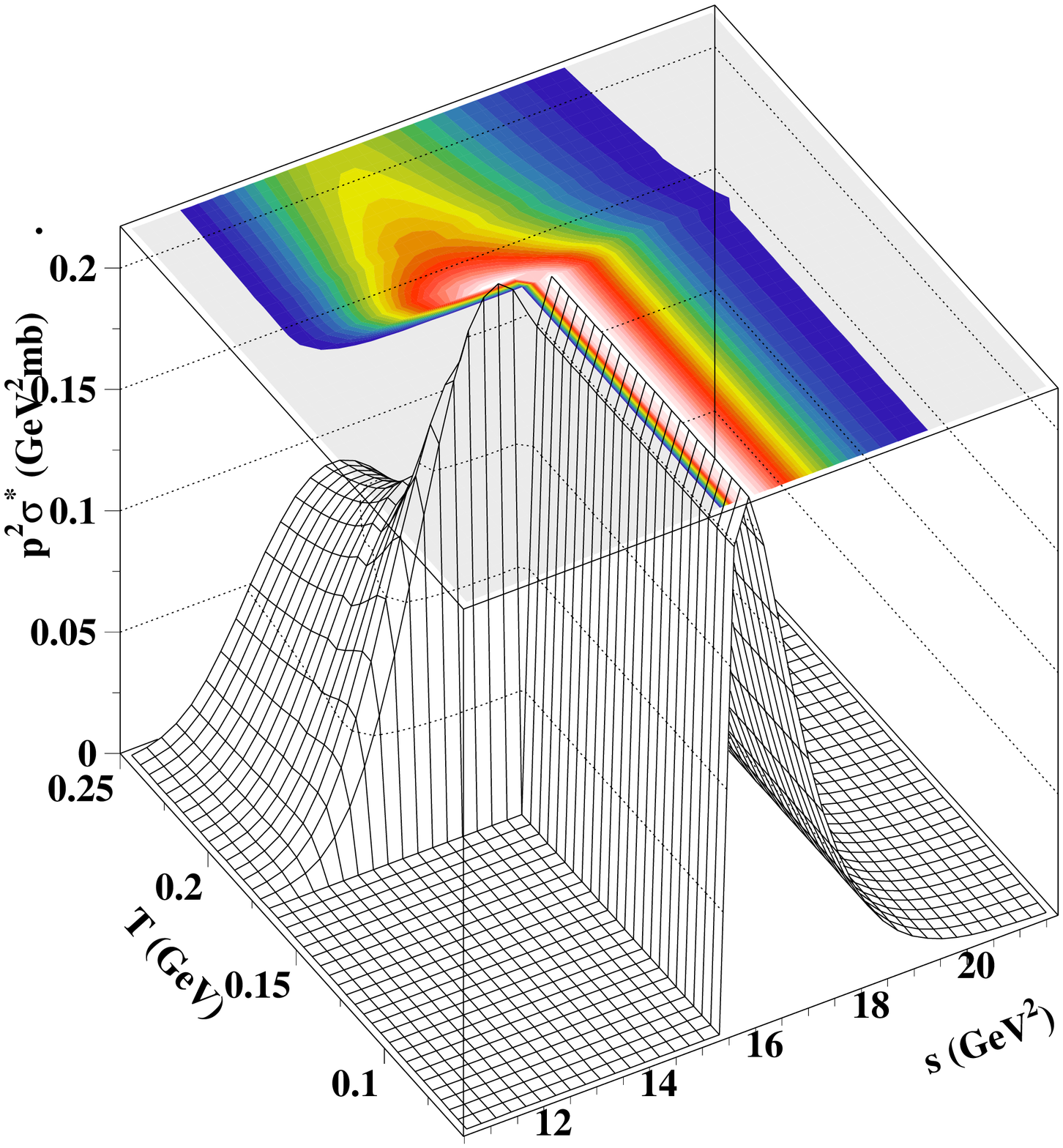}&
%}\hfill
%\parbox{9cm}{
%\includegraphics[width=6cm,height=5cm]{sigmaTPsiRhoNJL.eps}
\includegraphics[width=0.47\textwidth,height=0.35\textwidth]{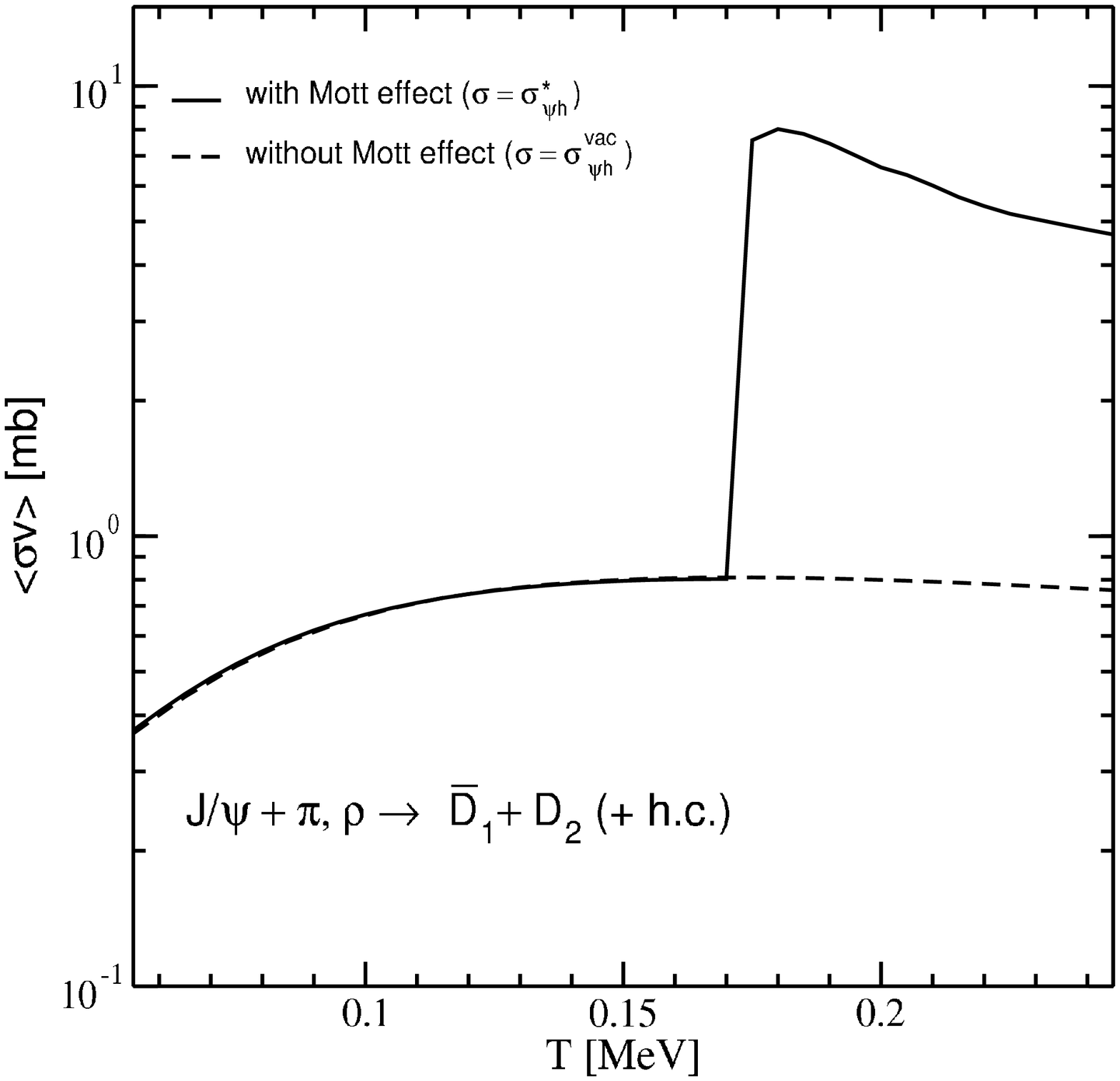}
\end{tabular}
%}
\caption{Left panel: energy ($s$) and temperature ($T$) 
dependence of the effective cross section ($\sigma^*$) for $J/\psi$ breakup 
by $\rho$-meson impact.
We display $p^2\sigma^*(s;T)$ for better visibility of the effective lowering 
of the breakup threshold when temperatures exceed the $D$-meson Mott 
temperature $T^{\rm Mott}\approx 172$ MeV;
right panel: temperature dependence of the thermally averaged $J/\psi$ 
breakup cross section in a $\pi$-$\rho$ meson gas; 
the calculation with vacuum $D$-mesons (dashed line) is compared to one 
with an in-medium broadening of the $D$-meson spectral function 
(due to the Mott effect at the chiral phase transition) which exhibits
a steplike enhancement (solid line) caused by the effective lowering 
of the breakup threshold; from Ref.~\cite{Blaschke:2002ww}.
\label{jpsidiss-medium}
}
\end{figure}

The energy dependent cross sections for heavy quarkonia
dissociation by hadron impact enable to evaluate the temperature- 
(and density-) dependent dissociation rates in hadronic matter.
Assuming the (short-distance) vertex functions not to be altered by 
the surrounding medium,
there remains the issue of  mass and widths changes
of open-charm hadrons with temperature and density. These, in particular,
imply modifications of the thresholds for the breakup 
processes~\cite{Grandchamp:2003uw,Sibirtsev:1999jr,Tsushima:2000cp,Burau:2000pn,Blaschke:2003ji,Fuchs:2004fh}.
In Fig.~\ref{jpsidiss-diagram} we show the diagrammatic representation of 
the quark-exchange contribution to the $J/\psi$ self-energy which develops 
an imaginary part (determining its width or inverse lifetime) due to the 
coupling to open-charm mesons. 
For the interaction vertex $U_{\rm ex}$ three different approaches have been 
discussed above and the corresponding vacuum cross sections are shown in 
Figs.~\ref{jpsimesondiss}-\ref{jpsidiss-chiLa}.
The theoretical basis for the discussion of quark exchange effects to the 
self-energy of heavy quarkonia in strongly correlated quark matter comes from
systematic cluster expansion techniques developed in the context of plasma
physics. For details, see the next subsection and Figs.~\ref{fig:cluster-ex},
\ref{fig:cluster-ex2} and \ref{fig:cluster-cluster}. 
Fig.~\ref{jpsidiss-medium} illustrates the effect of the spectral broadening
of $D$-mesons at the chiral phase transition due to the opening of the 
decay channel into their quark constituents (Mott effect) for temperatures
exceeding the $D$-meson Mott temperature $T^{\rm Mott}\approx 172$ MeV.
Due to an  effective lowering of the  $J/\psi$ breakup threshold the
temperature dependence of the thermally averaged $J/\psi$ 
breakup cross section in a $\pi$-$\rho$ meson gas exhibits a steplike 
enhancement~\cite{Blaschke:2002ww}.
This effect has been discussed as a possible mechanism underlying the
threshold-like anomalous suppression pattern of $J/\psi$'s observed by
the CERN NA50 experiment~\cite{Burau:2000pn,Blaschke:2000er} and should also
play a role in explaining the final state interactions of heavy quarkonia 
produced in the RHIC experiments.

\section{Perspectives}

In this lecture we have adapted a general thermodynamic Green function 
formalism developed in the context of nonrelativistic plasma physics for the
case of heavy quarkonia states in strongly correlated quark matter.
Besides the traditional folklore explanation of charmonium suppresion by 
(Debye-) screening of the strong interaction, we discuss further effects of
relevance when heavy quarkonia states propagate in a medium where strong 
correlations persist in the form of hadronic resonances.
These effects may be absorbed in the definition of a plasma Hamiltonian, which
was the main result of this contribution. This plasma Hamiltonian governs the 
in-medium modification of the bound state energy levels as well as the lowering
of the continuum edge which leads not only to the traditional Mott effect for
the dissociation of bound states in a plasma, but can also be applied to 
calculate the in-medium modification of quarkonium dissociation rates in a 
consistent way. 
A detailed recent review \cite{Rapp:2008tf} summarizes the phenomenological 
applications to heavy quarkonia production heavy-ion collision experiments at 
CERN and RHIC. 
Further developments of the presented approach shall include in particular 
applications to quarkonia production in dense baryonic matter such as envisaged
for the PANDA and CBM experiments at FAIR Darmstadt and possibly also for NICA 
at JINR Dubna.  

\section*{Acknowledgments}
This work has been supported in part by the Polish Ministry for Science
and Higher Education (MNiSW) under grants No. N N 202 0953 33 and 
No. N N 202 2318 37, and by the Russian Fund for Fundamental Investigations 
(RFFI) under grant No. 08-02-01003-a. 

\begin{footnotesize}

\end{footnotesize}
\end{document}